\documentclass[review,a4paper,12pt]{elsarticle}

\usepackage[margin=3cm]{geometry}

\usepackage{setspace}
\usepackage{amsmath,amssymb,amsfonts}
\usepackage{algorithmic}
\usepackage{graphicx}
\usepackage{textcomp}
\usepackage[table]{xcolor}
\usepackage{multicol}
\usepackage{cuted}
\usepackage[printonlyused, nolist]{acronym}
\usepackage{tabularx,booktabs}
\usepackage{verbatim}
\usepackage{import}
\usepackage{tikz}
\usepackage{float}
\def\checkmark{\tikz\fill[scale=0.4](0,.35) -- (.25,0) -- (1,.7) -- (.25,.15) -- cycle;} 

\usepackage{nomencl}
\usepackage{xstring}
\usepackage{etoolbox}
\usepackage{tocloft}  % safer control of list-like environments
\usepackage{xpatch}
\usepackage{eurosym}
\usepackage{flushend}
\usepackage{url}
\usepackage[normalem]{ulem} % para poder tachar cosas con \sout y que puedas verlas, para que decidas si tienen que seguir tachadas
\usepackage{hyperref} % hyperlink de referencias en el texto

\usepackage{multirow}
\usepackage{comment}

\newif\ifcomments\commentsfalse

\ifcomments
\usepackage{marginnote} 
\newcommand{\mymarginnote}[1]{\textcolor{blue}{\marginnote{#1}}}
\else
\newcommand{\mymarginnote}[1]{}
\fi

\ifcomments %% ESTO ES PARA MARCAR EN AZUL LAS REFERENCIAS NUEVAS, PON LA KEY DE LAS NUEVAS AHI DENTRO
\usepackage{xcolor,etoolbox}
\let\mybibitem\bibitem

\renewcommand{\bibitem}[1]{%
    \color{black}%
    \mybibitem{#1}%
}

\newcommand{\mycolor}[1]{%
\ifcomments \color{#1} \fi}
\else
\newcommand{\mycolor}[1]{}
\fi

\def\BibTeX{{\rm B\kern-.05em{\sc i\kern-.025em b}\kern-.08em
    T\kern-.1667em\lower.7ex\hbox{E}\kern-.125emX}}

\journal{EPSR}

\begin{document}

\title{Grid-Forming Converter DC-link Control Considering the Primary Energy Source\tnoteref{amnote}}

\author[Comillas]{Carlo De Paolis Robles}
\author[Comillas]{Andrés Tomás-Martín}
\author[Comillas]{Ignacio Egido}
\author[Comillas]{Aurelio García-Cerrada}

\affiliation[Comillas]{organization={Institute for Research in Technology (IIT), Comillas Pontifical University},%Department and Organization
            addressline={calle Alberto Aguilera, 23}, 
            postcode={28015},
            city={Madrid},
            country={Spain}}

\tnotetext[amnote]{\textcopyright~2026. This is the author's accepted manuscript
version of the article: C.~De Paolis Robles, A.~Tom\'as-Mart\'in, I.~Egido,
A.~Garc\'ia-Cerrada, ``Grid-Forming Converter DC-link Control Considering the
Primary Energy Source,'' \emph{Electric Power Systems Research}, Elsevier, 2026.
The final published version is available at
\href{https://doi.org/10.1016/j.epsr.2026.113887}{doi:10.1016/j.epsr.2026.113887}.
This manuscript version is made available under the CC~BY-NC-ND~4.0 license,
\url{https://creativecommons.org/licenses/by-nc-nd/4.0/}.}
\begin{frontmatter}
\begin{abstract}

The gradual substitution of conventional synchronous generators by converter-interfaced renewable energy sources raises concerns about the reduction of conventional inertia in electric power systems and the ensuing threat to their stability. In this regard, grid-forming voltage source converters have been proposed as a key solution to address this challenge. Although a growing body of literature addresses DC-link voltage regulation in grid-forming converters, most existing approaches implicitly assume an ideal and unconstrained DC power source. As a result, the dynamic response and operational limits of the primary energy source, which can critically shape the available DC-side power during transients, are rarely modeled or explicitly accounted for in the design of the DC-link control. This paper demonstrates that incorporating these aspects at the design stage reduces the risk of converter disconnection from the power grid under sudden power imbalances, while enhancing system resilience. A systematic methodology based on a Genetic Algorithm is proposed to tune the control parameters. The performance of the proposed control is validated by simulation using a detailed electromagnetic transient model.

\end{abstract}

\begin{keyword}
grid connectivity, grid-forming, primary energy source, DC-link, genetic algorithm.
\end{keyword}

\end{frontmatter}

\begin{acronym}[ROCOF] 
\acro{AC}{Alternating Current}
\acro{AS}{Ancillary Services}
\acro{AVR}{Automatic Voltage Regulator}
\acro{BESS}{Battery Energy Storage System}
\acro{CCT}{Critical Clearing Time}
\acro{COI}{Center of Inertia}
\acro{DC}{Direct Current}
\acro{DC-link}{Direct Current link}
\acro{DG}{Distributed Generation}
\acro{DSTATCOM}{Static Compensators connected in a Distributed system}
\acro{DVC}{DC-link Voltage Control}
\acro{dVOC}{Dispatchable Virtual Oscillator Control}
\acro{DVSC}{DC-link Voltage Synchronous Control}
\acro{EIR}{Inertia Emulation Controller}
\acro{EMT}{Electromagnetic Transient}
\acro{EMS}{Energy Management System}
\acro{ENTSO-E}{European Network of Transmission System Operators for Electricity}
\acro{eESM}{enhanced Electric Synchronous Machine}
\acro{ESM}{Electric Synchronous Machine}
\acro{ESS}{Energy Storage System}
\acro{EPC}{Electronic Power Converter}
\acro{FACTS}{devices for Flexible Alternating Current transmission systems}
\acro{FESS}{Flywheel Energy Storage Systems}
\acro{FFR}{Fast Frequency Response}
\acro{FOCV}{Fractional Open-Circuit-Voltage}
\acro{GA}{Genetic Algorithm}
\acro{GB}{Great Britain}
\acro{GFL}{Grid-Following}
\acro{GFL-VSC}{Grid-Following Voltage-Source Converter}
\acro{GFM}{Grid-Forming}
\acro{GFM-VSC}{Grid-Forming Voltage-Source Converter}
\acro{HVDC}{High Voltage Direct Current}
\acro{IAE}{Integral of Absolute Error}
\acro{IBR}{Inverter-Based Resources}
\acro{IDIk}{Inertia Distribution Index at bus k}
\acro{IIT}{Institute for Research in Technology}
\acro{IMDEA}{Madrid Institute for Advanced Studies}
\acro{IPFC}{Interline Power Flow Controller}
\acro{ITAE}{Integral Time Absolute Error}
\acro{IR}{Inertial Response}
\acro{MFB}{Most Flexible Bus}
\acro{ML}{Machine Learning}
\acro{MMC}{Modular Multilevel Converter}
\acro{MPPT}{Maximum Power Point Tracking}
\acro{NESO}{National Energy System Operator}
\acro{NMPC}{Nonlinear Model Predictive Control}
\acro{NRPG}{Northern Regional Power Grid}
\acro{nSG}{Non-Synchronous Generator}
\acro{OT}{Optimization Techniques}
\acro{OWF}{Offshore Wind Farm}
\acro{PDF}{Probability Density Function}
\acro{PES}{Primary Energy Source}
\acro{PFR}{Primary Frequency Regulation}
\acro{Ph.D.}{Doctor of Philosophy}
\acro{PD}{Proportional-Derivative}
\acro{PI}{Proportional-Integral}
\acro{PID}{Proportional-Integral-Derivative}
\acro{PLL}{Phase-Locked Loop}
\acro{PMSG}{Permanent Magnet Synchronous Generator}
\acro{PMU}{Phasor Measurement Unit}
\acro{PPM}{Power Park Module}
\acro{PSO}{Particle Swarm Optimization}
\acro{PSS}{Power System Stabilizers}
\acro{PV}{Photovoltaic}
\acro{RAGFM}{Resource-Aware Grid-Forming}
\acro{RES}{Renewable Energy Source}
\acro{ROCOF}{Rate of Change of Frequency}
\acro{RMS}{Root Mean Square}
\acro{SOC}{State of Charge}
\acro{S1}{Center Oscillation Index}
\acro{S2}{Center of Frequency Index}
\acro{SG}{Synchronous Generator}
\acro{ST}{settling time}
\acro{STATCOM}{Static Synchronous Compensators}
\acro{UC}{ultracapacitor}
\acro{UK}{United Kingdom}
\acro{VCVSC}{Voltage Control Voltage Source Converter}
\acro{VFLEXP}{Vector-Based Flexible-Complexity Power System Tool}
\acro{VI}{Virtual Inertia}
\acro{VIMP}{Virtual Impedance}
\acro{VqFF}{q-axis voltage feedforward}
\acro{VQ-VSC}{Voltage-Controlled Voltage Source Converter}
\acro{VSC}{Voltage Source Converter}
\acro{VSG}{Virtual Synchronous Generator}
\acro{VSM}{Virtual Synchronous Machine}
\acro{WPP}{Wind Power Plant}
\acro{WTG}{Wind Turbine Generator}
\end{acronym}

\patchcmd{\thenomenclature}
  {\leftmargin\labelwidth}
  {\leftmargin\labelwidth\itemindent 1em }
  {}{}
\newcommand{\nomenclheader}[1]{%
  \item[\hspace*{-\itemindent}\normalfont\bfseries#1]}
\renewcommand\nomgroup[1]{%
  \IfStrEqCase{#1}{
   %{S}{\nomenclheader{Sets and indexes:}}
   %{P}{\nomenclheader{Parameters:}}
   %{V}{\nomenclheader{Variables:}}
   %{F}{\nomenclheader{ROCOF UC:}}
   %{S}{\nomenclheader{Standard UC:}}
   {A}{\nomenclheader{Genetic Algorithm:}}
  }%
}

\makenomenclature

\printnomenclature
\nomenclature[A]{$e_p(t)$}{Instantaneous absolute error in output power}
\nomenclature[A]{$e_{u^2}(t)$}{Instantaneous absolute error in squared DC voltage}
\nomenclature[A]{$\text{IAE}_p$}{Integral of Absolute Error for output power}
\nomenclature[A]{$\text{IAE}_{u^2}$}{Integral of Absolute Error for squared DC-voltage}
\nomenclature[A]{$J$}{Fitness function score to be minimized}
\nomenclature[A]{$lb$}{Parameter lower bound}
\nomenclature[A]{$\text{penalty}$}{Penalty term added for constraint violation}
\nomenclature[A]{$p_{\text{load}}$}{Load setpoint}
\nomenclature[A]{$p_{\text{out}}$}{Converter output power}
\nomenclature[A]{$\Delta p_{\text{out}}$}{Incremental converter output power}
\nomenclature[A]{$p_{\text{out-sp}}$}{Converter output power setpoint}
\nomenclature[A]{$\Delta p_{\text{out-sp}}$}{Incremental converter output power setpoint correction (output of PD-out)}
\nomenclature[A]{$p_{\text{pes}}$}{Primary Energy Source power}
\nomenclature[A]{$p_{\text{pes-sp}}$}{Primary Energy Source power setpoint}
\nomenclature[A]{$T$}{Simulation time horizon for error integration}
\nomenclature[A]{$ub$}{Parameter upper bound}
\nomenclature[A]{$u_{\text{dc}}^2$}{Squared DC-link voltage (controlled variable)}
\nomenclature[A]{$\Delta u_{\text{dc}}^2$}{Squared DC-link voltage error between the setpoint and the actual value}
\nomenclature[A]{$u_{\text{dc-max}}$}{Upper safety limit of the DC-link voltage}
\nomenclature[A]{$u_{\text{dc-min}}$}{Lower safety limit of the DC-link voltage}
\nomenclature[A]{$u_{\text{dc-sp}}^2$}{Squared DC-link voltage setpoint}
\nomenclature[A]{$w_1$}{Weighting factor for power error}
\nomenclature[A]{$w_2$}{Weighting factor for squared DC-link voltage error}

\section{Introduction}\label{sec.1_introduction}

A significant transformation in power systems is underway due to the increased penetration of \acp{RES} interfaced through \acp{EPC} and the retirement of conventional \acp{SG}, leading to a marked reduction in rotational inertia~\cite{shazon2022frequency}. Low-inertia systems exhibit larger frequency excursions after disturbances because physical inertia reduces the \ac{ROCOF} before primary control reacts~\cite{saha2023impact}. While \acp{EPC} can deliver fast frequency support, it remains uncertain whether \ac{FFR} alone can fully compensate for this loss~\cite{dorfler2023control}. To address this, the \ac{ENTSO-E} proposed in 2023 a set of mitigation measures~\cite{ProjectInertiaPhase}, distinguishing Foundational Measures, which include \acp{VSC} with \ac{GFM} capability able to emulate virtual inertia, from Enhanced Response Measures, based on \ac{GFL} converters that provide rapid support but cannot respond instantaneously due to their reliance on the detection of grid voltage~\cite{li2022revisiting}. Within this framework, the development and deployment of \acp{GFM-VSC} emerge as a key solution for maintaining adequate frequency stability in future low-inertia systems~\cite{zhang2021grid}.

Recent studies on frequency stability in converter-dominated systems model the \ac{DC-link} side with varying degrees of detail. Many system-level analyses still assume an ideal \ac{DC-link} and an instantly available primary energy source, thereby neglecting \ac{DC}-side dynamics and limits when studying inertia and \ac{ROCOF} behavior~\cite{mahmood2024evaluating,ducoin2024analytical}. Some even claim that the \ac{DC-link} has little influence on frequency regulation as long as neither the source nor the converter saturates~\cite{kenyon2023interactive}. By contrast, a growing line of research explicitly accounts for \ac{DC-link} capacitor dynamics and develops dedicated voltage-control strategies, recognizing that robust frequency regulation and transient stability depend on proper \ac{DC-link} voltage management.

Some authors study \ac{DC-link} dynamics from a system-level stability viewpoint, embedding grid-forming converter models with explicit \ac{DC-link} dynamics in large benchmark networks (e.g., IEEE 9--68 bus test systems) to assess impacts on closed-loop stability and grid interactions. Reference~\cite{samanta2021stability} shows that matching-controlled \acp{GFM-VSC} possess inherent \ac{DC-link} voltage stability under \ac{DC}-side current saturation, unlike other \ac{GFM} strategies, extending the framework of~\cite{tayyebi2020frequency}. A follow-up study~\cite{samanta2022fast} addresses this limitation with a sliding-mode controller that stabilizes the \ac{DC-link} voltage and provides quantifiable \ac{FFR}, though without ensuring optimality or explicitly handling input constraints. More recently, \mymarginnote{R2}\mycolor{blue}the work in\mycolor{black}~\cite{samanta2023nonlinear} introduces an \ac{NMPC} formulation that incorporates AC/\ac{DC} current limits and control-input bounds, enabling optimal and tunable \ac{FFR}. Complementarily, \mycolor{blue}the authors in\mycolor{black}~\cite{karunaratne2023nonlinear} propose a decentralized nonlinear backstepping controller that regulates the \ac{DC-link} voltage via active-power adaptation, offering Lyapunov-certified stability with low computational effort. All these references assume a fast \ac{PES} represented as an instantaneous current-limited source.

Several authors address \ac{DC}-side control from a device-level perspective, analyzing how \ac{DC-link} voltage dynamics couple with grid-forming loops. These works emphasize converter-level modeling and controller design (structure, gains, stability margins, and transients), typically via small-signal analysis and time-domain simulations, and sometimes with laboratory prototype validation. Reference~\cite{shen2023transient} shows how these dynamics influence the transient stability of \acp{VSG} and classifies \ac{DC-link} controllers according to the \ac{AC}-side variable they modify, either the active-power reference or the internal frequency, while noting that other structures are possible depending on the \ac{GFM} architecture.

A first group of references regulates the \ac{DC-link} voltage by adjusting the converter's active power output setpoint. Reference~\cite{girona2024resource} embeds the \ac{DC-link} voltage into the synchronization loop to guarantee regulation while preserving \ac{GFM} behavior. Reference~\cite{xu2025stability} places a stronger emphasis on transient stability, proposing a dedicated \ac{PI} controller that forces the \ac{DC-link} capacitor to temporarily absorb power during severe voltage or angle disturbances at the point of connection. This controlled energy absorption effectively reduces the virtual inertia provided by the converter and \mymarginnote{R2}\mycolor{blue}enhances \mycolor{black}damping, \mycolor{blue}thereby suppressing \mycolor{black}\ac{DC-link} overvoltage and \mycolor{blue}improving \mycolor{black}the converter's transient stability margin. Reference~\cite{luo2023design} shows that a \ac{PI}-based controller acting on the squared voltage may destabilize a droop-controlled \ac{GFM} inverter during undervoltage events if poorly tuned. All three references use an ideal constant-power \ac{PES} model without explicitly considering its dynamic characteristics.

A second group of references regulates the \ac{DC-link} voltage by adjusting the converter's internal frequency. Reference~\cite{zhao2023small} uses a frequency-modulation scheme with a \ac{VqFF} loop to add damping and improve voltage regulation. Reference~\cite{qin2024novel} introduces a compensation-based strategy for full-scale \acp{WTG}, where a lead--lag compensator counteracts the negative damping from large inertia gains, enabling higher inertial coefficients without destabilizing the \ac{DC-link}. In~\cite{tian2023two}, deviations of the squared \ac{DC-link} voltage directly modulate the internal frequency, allowing inertia emulation without auxiliary storage and preserving \ac{GFM} behavior even in weak grids. As in the previous group, these studies assume an instantaneous or constant-power \ac{PES}, neglecting its dynamics.

Finally, other \ac{DC-link} control strategies target different variables within the \ac{GFM} structure. Reference~\cite{peng2022transient} enhances transient stability by acting on the virtual rotor angle inferred from the \ac{DC-link} dynamics, injecting damping to suppress voltage excursions. Other references regulate the \ac{DC-link} via a current-controlled ultracapacitor connected through a \ac{DC}/\ac{DC} converter. In~\cite{gross2022energy}, a cascaded \ac{DC}-side controller manages the \ac{DC-link} voltage exclusively through the ultracapacitor branch, with an \ac{EMS} ensuring safe \ac{UC} operation during active-power support. Reference~\cite{kryonidis2023use} provides experimental validation of this architecture for inertial response, showing that ultracapacitors can deliver fast frequency support without compromising \ac{DC-link} stability. These three references model the \ac{PES} as a constant power injection, without incorporating any dynamic behavior.

Table~\ref{tab_articles_comparison} compiles the most relevant references discussed. For each work, it identifies whether a dedicated \ac{DC-link} voltage-control strategy is considered and whether this control action coordinates appropriately with a \ac{GFM-VSC} that provides inertia emulation and \ac{PFR}. The table also indicates whether the proposed control restores the \ac{DC-link} voltage to its nominal value following a grid-side power disturbance, and it reports the control design methodology adopted in each study. Finally, it clarifies whether upstream \ac{PES} dynamics on the \ac{DC} side are modeled and whether the \ac{DC-link} control explicitly accounts for them to enable grid connectivity while providing frequency support.

\begin{table*}[!ht]
\caption{Comparison of the most relevant references from the Introduction on \ac{DC-link} methods}
\centering
\footnotesize
\begin{tabular}{c c c c c c c}
\toprule
\textbf{Reference} &
\textbf{IE} &
\textbf{PFR} &
\textbf{DCNV} &
\textbf{Design method} &
\textbf{PESd} &
\textbf{PESc} \\
\midrule
\cite{samanta2023nonlinear}      & \checkmark & \checkmark & \checkmark & \ac{NMPC}      & \checkmark$^{*}$ &  \\
\cite{karunaratne2023nonlinear}  & \checkmark & \checkmark & \checkmark & Backstepping    & \checkmark$^{*}$ &  \\
\cite{shen2023transient}         & \checkmark & \checkmark & \checkmark & Lyapunov &  &  \\
\cite{girona2024resource}        & \checkmark & \checkmark &  & Analytical SG swing eqs. &  & \checkmark \\
\cite{xu2025stability}           & \checkmark & \checkmark & \checkmark & Lyapunov &  &  \\
\cite{luo2023design}             &  & \checkmark & \checkmark & Bifurcation-based &  &  \\
\cite{zhao2023small} &  &  &  & Small-signal and pole-analysis &  &  \\
\cite{qin2024novel}  & \checkmark & \checkmark & \checkmark 
& Three methods$^{\square}$ 
&  &  \\
\cite{tian2023two}               & \checkmark & \checkmark & \checkmark & PI-based second-order design &  &  \\
\cite{peng2022transient}         & \checkmark & \checkmark & \checkmark & Passivity-based / Hamiltonian &  &  \\
\cite{gross2022energy}           &  &  & \checkmark & Cascaded PI &  &  \\
\cite{kryonidis2023use}   & \checkmark &  & \checkmark & Cascaded PI &  &  \\

\textbf{This \mbox{article}}            & \checkmark & \checkmark &\checkmark & \ac{GA} & \checkmark & \checkmark  \\

\bottomrule
\end{tabular}
\begin{flushleft}
\checkmark$^{*}$ \ac{PES} dynamic represented as a first-order transfer function with 1~ms time constant \\
$^{\square}$Dynamic DC-link control + damping compensation + lead--lag inertial compensation \\
IE: Inertia Emulation, PFR: Primary Frequency Regulation, DCNV: DC-link Controlled to Nominal Value, PESd: Primary Energy Source Dynamic Consideration, PESc: Primary Energy Source Control
\end{flushleft}
\label{tab_articles_comparison}
\end{table*}

This paper investigates how a \ac{GFM-VSC} can jointly regulate its \ac{DC-link} voltage and the power supplied to the grid to remain connected after power disturbances, explicitly modeling \ac{PES} dynamics rather than assuming an instantaneous or constant-power source. The proposed approach controls both the energy extracted from the \ac{PES} and the converter active-power output setpoint similarly to \mymarginnote{R2}\mycolor{blue}the approach in\mycolor{black}~\cite{girona2024resource}, which, to the authors' knowledge, is the only existing work that acts on both sides simultaneously. This article makes the following contributions: (i) a \ac{DC-link} voltage control strategy for \acp{GFM-VSC} that provides a fast response to \ac{AC}-side power imbalances by coordinating the primary energy source and converter power control, thereby enforcing strict \ac{DC-link} voltage limits and avoiding converter disconnections; and (ii) a Genetic Algorithm-based parameter-tuning methodology that optimizes the disturbance step-response performance, first developed on a simplified model and then validated using a detailed \ac{EMT} model.

The paper is organized as follows. Section~\ref{sec.2_problem_and_case_study} describes the problem in detail and presents the proposed approach. The control configuration for the \ac{DC-link} voltage of a \ac{GFM-VSC} is explained. It also describes the case study to be used in the following sections. Section~\ref{sec.3_methodology} explains the methodology \mymarginnote{R2}\mycolor{blue}used for controller tuning\mycolor{black}. Section~\ref{sec.4_detailed_model_test} \mycolor{blue}verifies the performance of the parameter-tuned controller using a detailed \ac{EMT} model in \mycolor{black}the case study. Section~\ref{sec.5_conclusions} concludes the paper and suggests future work.

\section{Description of the Problem and Case Study}\label{sec.2_problem_and_case_study}

To ensure the stable grid connectivity of \acp{GFM-VSC}, it is critical to coordinate the interactions between the \ac{PES}, the \ac{AC} output power, and the \ac{DC-link} energy balance. This section first establishes the modeling hypotheses and system abstraction required to capture these dynamics, characterizing the simplified representations of the network, the converter, and the primary energy source. Building upon this analytical framework, a novel \ac{DC-link} supervisory control philosophy is proposed, specifically designed to manage the \ac{GFM-VSC} and \ac{PES} interactions within the relevant operational time scales.

\begin{figure}[h]
    \centering
    \includegraphics[width = \linewidth]{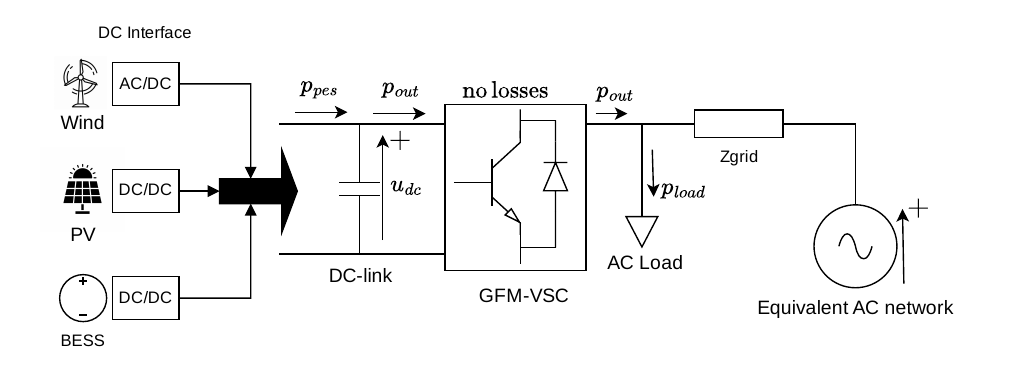} 
    \caption{Case study: \ac{GFM-VSC} with \ac{DC}-side constraints connected to an \ac{AC} grid}
    \label{fig_case_study_system}
\end{figure}

The case study, illustrated in Figure~\ref{fig_case_study_system}, comprises a \ac{PES} (representing wind, solar, or \ac{BESS}), a \ac{DC-link} capacitor, and a \ac{GFM-VSC} interface. The \ac{AC} grid is modeled as an equivalent \ac{AC} voltage source behind a predominantly inductive grid impedance ($Z_{grid}$).

\subsection{Modeling Hypotheses and System Abstraction}\label{subsec.2_1_modeling_hypothesis}

To ensure the tractability of the optimization problem while maintaining physical validity, the following modeling assumptions are explained in this subsection.

\subsubsection{Network Dynamics}\label{subsubsec.2_1_1_network_dynamics}

The AC-grid voltage is assumed to be constant, and the grid impedance $Z_{grid}$ is characterized as predominantly inductive ($X \gg R$). Under these conditions, the power exchange is governed by the power-angle relationship:

\begin{equation}
P \approx \frac{V_1 V_2}{X} \sin(\delta) \approx \frac{V_1 V_2}{X} \delta
\label{eq.power_exchange}
\end{equation}

where $\delta$ represents the phase shift between the converter's internal voltage ($V_1$) and the grid voltage ($V_2$). Following a frequency disturbance, the phase angle $\delta$ deviates as a function of the frequency integral, expressed as $\Delta \delta = \int 2\pi \Delta f \, dt$, in line with references~\cite{entsoe2025_phaseII_news} and~\cite{klaes2025immunity}. In compliance with the \ac{ENTSO-E} technical requirements for grid-forming capability, \ac{GFM} units must act as a voltage source behind an internal impedance, maintaining their internal voltage amplitude and phase angle constant at the inception of a network disturbance~\cite{entsoe2025_phaseII_news}. Specifically, the synthetic inertia requirement implies that a \ac{PPM} must provide an active power change $\Delta P$ proportional to the \ac{ROCOF}, such that:

\begin{equation}
\Delta P = T_{M,PPM} \cdot \frac{df/dt}{f_{rated}}
\label{eq.rocof_inertia_entsoe}
\end{equation}

where $T_{M,PPM} = 2H$. This expression implies that grid power imbalances, such as a load increase or generator loss, are directly reflected in the converter's synchronization dynamics.

\subsubsection{Converter and Grid-Forming Dynamics}\label{subsubsec.2_1_2_converter_dynamics}

A cascaded control structure for a \ac{GFM} \ac{VSM} connected via an $LC$ or $LCL$ filter constitutes a nonlinear system involving several coupled dynamics~\cite{avila2026impact}. The main dynamics of the cascaded control loops, modeled in a synchronous $d$--$q$ reference frame, are expressed as follows:

\paragraph{Synchronization and Converter Dynamics}\label{parag.2_1_2_1_dynamics}
\begin{itemize}
    \item Inner Current Loop: The inner loop controls the filter current ($i_f$) to track its reference ($i_f^*$). Including cross-coupling decoupling and capacitor voltage feedforward, the closed-loop transfer function is:
    \begin{equation}
    G_c(s) = \frac{i_{f_{d,q}}(s)}{i_{f_{d,q}}^*(s)} = \frac{K_{pc} s + K_{ic}}{L_f s^2 + (R_f + K_{pc})s + K_{ic}}
    \label{eq.current_loop_tf}
    \end{equation}

    \item Inner Voltage Loop: The voltage loop regulates the filter capacitor voltage ($v_c$). Assuming ideal current tracking ($i_f \approx i_f^*$), the closed-loop response is:
    \begin{equation}
    G_v(s) = \frac{v_{c_{d,q}}(s)}{v_{c_{d,q}}^*(s)} = \frac{K_{pv} s + K_{iv}}{C_f s^2 + K_{pv} s + K_{iv}}
    \label{eq.voltage_loop_tf}
    \end{equation}

    \item Reactive Power-Voltage ($Q-V$) Droop: Regulates the voltage magnitude reference ($V_{ref}^*$) based on the measured reactive power ($Q_{meas}$):
    \begin{equation}
    V_{ref}^*(s) = V_{nom} - D_q \left( Q_{meas}(s) - Q_{ref}(s) \right)
    \label{eq.reactive_power_droop}
    \end{equation}

    \item Active Power-Angle \mycolor{blue}($P-\theta$) \mycolor{black}\ac{VSM} Outer Loop: Emulates virtual inertia ($H$) and damping ($D$) via the swing equation:
    \begin{equation}
    2H s \Delta \omega(s) = P_{ref}(s) - P_{meas}(s) - D \Delta \omega(s)
    \label{eq.swing_equation}
    \end{equation}
    The phase angle is obtained by integrating the frequency deviation:
    \begin{equation}
    \Delta \theta(s) = \frac{\Delta \omega(s)}{s} \implies \frac{\Delta \theta(s)}{P_{ref}(s) - P_{meas}(s)} = \frac{1}{s(2Hs + D)}
    \label{eq.vsm_angle_tf}
    \end{equation}
\end{itemize}

\paragraph{Simplification Hypotheses}\label{parag.2_1_2_2_simp_hypot}

To reduce the dynamics of a \ac{GFM-VSC} to a proposed first-order equivalent model $1/(T_c s + 1)$ (see Figure~\ref{fig_block_diagram_1}), three justified assumptions are sequentially applied:

\begin{enumerate}
    \item Hypothesis 1---$P$--$\theta$ and $Q$--$V$ Decoupling: Building upon the network assumptions in Section~\ref{subsubsec.2_1_1_network_dynamics}, the predominantly inductive grid ($X \gg R$) ensures that active power transfer is tightly coupled to the power angle ($\delta$) and decoupled from voltage magnitudes. While the grid voltage magnitude $V_2$ is assumed constant (Section~\ref{subsubsec.2_1_1_network_dynamics}), the converter's internal voltage magnitude $V_1$ is regulated by the $Q$--$V$ droop loop (\mymarginnote{R2}\mycolor{blue}Equation\mycolor{black}~\eqref{eq.reactive_power_droop}) and, in compliance with the \ac{ENTSO-E} grid-forming requirements, is maintained close to its nominal value during active power transients. Under these conditions, this relationship is linearized as $\Delta P \approx K_s \Delta \theta$, where $K_s$ is the synchronizing power coefficient derived from \mycolor{blue}Equation\mycolor{black}~\eqref{eq.power_exchange}. This justifies neglecting $Q$--$V$ loop dynamics during active power tracking analysis.
    
    \item Hypothesis 2---Time-Scale Separation: Following singular perturbation theory~\cite{rouco2013selective}, the cascaded structure exhibits a clear time-scale separation between the inner electromagnetic loops (current and voltage, \mymarginnote{R2}\mycolor{blue}Equation\mycolor{black}~\eqref{eq.current_loop_tf}--\eqref{eq.voltage_loop_tf}) and the outer electromechanical loop (the \ac{VSM} active power--angle loop in \mycolor{blue}Equation\mycolor{black}~\eqref{eq.swing_equation}), whose role is to emulate virtual inertia and damping and to set the converter's phase angle. Since the inner loops are orders of magnitude faster than the outer \ac{VSM} dynamics, from the perspective of the outer loop their transients decay instantaneously. Mathematically, this allows the inner loops to be evaluated at their steady-state limit ($s \to 0$), yielding unity gains ($G_c(0) \approx 1, G_v(0) \approx 1$).
    
    \item Hypothesis 3---Dominant Pole Approximation: Applying the previous hypotheses, the plant collapses into the closed-loop active power response of the outer \ac{VSM} loop:
    \begin{equation}
    G_{vsm\_p}(s) = \frac{P_{meas}(s)}{P_{ref}(s)} = \frac{K_s}{2H s^2 + D s + K_s} = \frac{\omega_n^2}{s^2 + 2\zeta\omega_n s + \omega_n^2}
    \label{eq.vsm_canonical_form}
    \end{equation}
    where $\omega_n = \sqrt{K_s / 2H}$ and $\zeta = D / (2 \sqrt{2H K_s})$. Tuning $D$ for a critically damped or overdamped response ($\zeta \ge 1$) allows the active power injection to be accurately modeled by a first-order lag with time constant $T_c$, capturing the dominant time constant of the system.
\end{enumerate}

Therefore, the \ac{GFM-VSC} is simplified to a first-order transfer function $1/(T_c s + 1)$ with a time constant $T_{c} \approx 100$~ms, which is a value taken from experimental results in~\cite{tomas2026improvements}. This abstraction captures the dynamic response of the \ac{GFM-VSC} 
as an equivalent mapping from the commanded converter output power setpoint to the actual converter output power, which determines the influence of the \ac{AC} side on the \ac{DC-link} power balance, as represented later in Section~\ref{subsec_2_2_philosophy}.

\subsubsection{Primary Energy Source Dynamics}\label{subsubsec.2_1_3_pes}

To assess the impact of the assumed \ac{PES} dynamics, the \ac{PES} was modeled with a first-order transfer function with a time constant $T_{PES}$. This model captures the physical limitations and power delivery delays of the energy source interface. Although this representation does not account for all higher-order effects, a first-order model is widely adopted for system-level stability and control studies, as evidenced in~\cite{samanta2021stability,tayyebi2020frequency, samanta2022fast, samanta2023nonlinear, karunaratne2023nonlinear}.

The validity of this simplification rests on the principle of time-scale separation, which holds that the system's transient response is primarily determined by its slowest time constant~\cite{rouco2013selective}. In this context, the overall response speed of the power delivery has a greater influence on \ac{DC-link} stability than the specific mathematical order of the underlying internal dynamics.

It should be emphasized that the proposed control design and optimization framework is not inherently limited to first-order \ac{PES} approximations. The methodology is sufficiently flexible to be readily extended to higher-order models, ensuring its applicability across a wide range of \ac{PES} configurations.

\subsubsection{DC-link Energy Balance}\label{subsubsec.2_1_4_dc_link}

The \ac{DC-link} is modeled as an equivalent capacitance $C$. Acting as an energy buffer, this capacitance can be characterized by an equivalent inertia constant, as proposed in references~\cite{yang2022internal} and~\cite{shafique2025dc}. This approach allows us to map the physical capacitance (in Farads) to a time constant (in seconds), as utilized in Section~\ref{subsec.3_4_1_base_case}.

\paragraph{Energy characterization}\label{parag.2_1_4_1_energy_charac}
Building upon the fundamental concepts presented in~\cite{yang2022internal} and~\cite{shafique2025dc}, the total energy stored in the capacitive elements of a converter, denoted as $W_{\text{total}}$, is a function of the total capacitance $C_{\text{total}}$ and the operating \ac{DC} voltage $V_{\text{dc}}$: 

\begin{equation}
W_{\text{total}} = \frac{1}{2} C_{\text{total}} V_{\text{dc}}^2
\label{eq.capacitor_energy_storage}
\end{equation}

As highlighted in reference~\cite{yang2022internal}, $C_{\text{total}}$ represents the lumped equivalent capacitance of the system. This energy serves as a physical buffer to compensate for instantaneous active power imbalances between the \ac{DC} and \ac{AC} sides.

\paragraph{DC Inertia Constant ($H_{\text{DC}}$) Formulation}
Based on the dynamic power balance principles detailed in~\cite{yang2022internal} and~\cite{shafique2025dc}, the energy stored in the \ac{DC-link} capacitor, $E_{\text{stored}}$, reflects the time integral of the active power mismatch between the \ac{DC}-side input ($P_{\text{DC}}$) and the \ac{AC}-side output ($P_{\text{AC}}$):

\begin{equation}
E_{\text{stored}} = \int \left( P_{\text{DC}} - P_{\text{AC}} \right) dt = \frac{1}{2} C_{\text{total}} V_{\text{dc}}^2
\label{eq.capacitor_stored_energy}
\end{equation}

where $E_{\text{stored}}$ is the total energy stored in the capacitor, $P_{\text{DC}}$ is the \ac{DC}-side input active power, $P_{\text{AC}}$ is the \ac{AC}-side output active power, $C_{\text{total}}$ is the aforementioned total \ac{DC-link} capacitance, and $V_{\text{dc}}$ is the \ac{DC} voltage.

To express this physical storage as a time constant (i.e., the \ac{DC} inertia constant $H_{\text{DC}}$), the stored energy is normalized by the nominal apparent power of the converter, $S_N$. This establishes the equivalence between the actual capacitance ($C_{\text{F}}$, in Farads) and the time constant ($C_{\text{s}}$, in seconds):

\begin{equation}
E_{\text{stored}} = S_N \cdot C_{\text{s}} = \frac{1}{2} C_{\text{F}} V_{\text{dc,nom}}^2
\label{eq.capacitor_mF_to_s}
\end{equation}

where $S_N$ is the nominal converter power (in VA), $C_{\text{s}}$ is the equivalent \ac{DC-link} capacitance expressed in seconds (defined as the \ac{DC} inertia constant $H_{\text{DC}}$ in~\cite{shafique2025dc}), and $V_{\text{dc,nom}}$ is the nominal \ac{DC} voltage.

Thus, the \ac{DC} inertia constant $H_{\text{DC}}$ is mathematically expressed as follows:

\begin{equation}
H_{\text{DC}} (\text{s}) = C_{\text{s}} = \frac{C_{\text{F}} V_{\text{dc,nom}}^2}{2 S_N} 
\label{eq.DC_inertia_constant}
\end{equation}

Applying this generalized formulation to the parameters provided in~\cite{yang2022internal} (1~GW, 640~kV, and 412 series modules, 6 capacitors of 11~mF each), the resulting \ac{DC} inertia constant is $H_{\text{DC}} \approx 30$~ms. This benchmark is highly consistent with the parameter values applied in Section~\ref{subsec.3_4_1_base_case}, validating the proposed \ac{DC-link} buffer sizing.

\subsection{Proposed DC-link Supervisory Control Philosophy}\label{subsec_2_2_philosophy}

\begin{figure}[h]
    \centering
    \includegraphics[width = \linewidth]{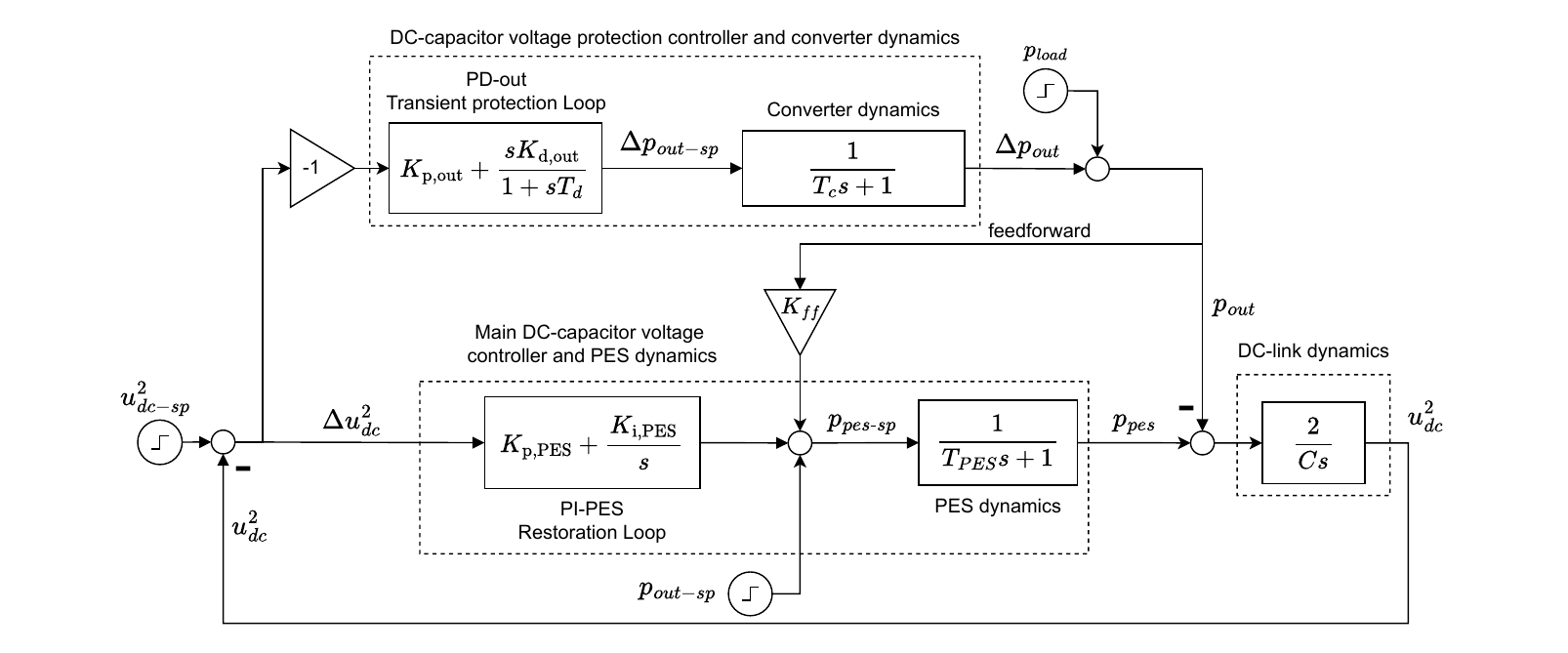} 
    \caption{Reduced-order control block diagram representing the supervisory layer and system dynamics}
    \label{fig_block_diagram_1}
    
\end{figure}

Although the \ac{PES}-side controller and the converter-side controller, shown in Figure~\ref{fig_block_diagram_1}, are both driven by the same feedback signal $\Delta u^{2}_{\text{dc}}$, their coordination is ensured by a deliberate separation of roles and time scales:

\begin{enumerate}
    \item Restoration Loop (\textit{PI-PES}): This controller handles long-term energy balance, restoring $u_{\text{dc}}^2$ to its nominal value by acting on the \ac{PES} setpoint ($p_{\text{pes-sp}}$), owing to the integral action. The derivative component is intentionally excluded to avoid excessive setpoint excursions to the \ac{PES} and prevent its power limits from being exceeded. Finally, this setpoint ($p_{\text{pes-sp}}$) is processed by the inherent \ac{PES} dynamics, where the response is governed by the time constant \mymarginnote{R2}\mycolor{blue}$T_{PES}$\mycolor{black}, ultimately determining the actual power delivered by the \ac{PES} ($p_{\text{pes}}$).

    \item Transient Protection Loop (\textit{PD-out}): This controller acts on the \ac{GFM} output power reference to prevent the \ac{DC-link} voltage from exceeding safety limits $[u_{\text{dc-min}}, u_{\text{dc-max}}] = [0.7, 1.15]$~p.u. It is implemented as a \ac{PD} action and intentionally excludes integral action. The inverting gain at the controller input ensures that the output of \textit{PD-out}, $\Delta p_{\text{out-sp}}$, reduces the converter setpoint during \ac{DC-link} voltage drops. This setpoint correction ($\Delta p_{\text{out-sp}}$) is then processed by the converter dynamics, where the response is governed by the time constant $T_c$, ultimately determining the resulting variation in the converter output power ($\Delta p_{\text{out}}$). The total converter output power ($p_{\text{out}}$) delivered to the grid then results from combining this variation $\Delta p_{\text{out}}$ with the baseline setpoint associated with the system load demand $p_{\text{load}}$. The proposed control strategy does not affect the steady-state behavior of the \ac{GFM-VSC} on the \ac{AC} side, as is normally required for \ac{GFM} devices. The \ac{DC-link} control strategy takes over power transients when $u_{\text{dc}}^2$ changes.
\end{enumerate}

Building on Equation~\eqref{eq.capacitor_stored_energy}, with $P_{\text{AC}} = p_{\text{out}}$ and $P_{\text{DC}} = p_{\text{pes}}$, the time integral of the power imbalance yields the squared \ac{DC-link} voltage $u_{\text{dc}}^2$, as represented by the \ac{DC-link} dynamics block in Figure~\ref{fig_block_diagram_1}. This design prevents both loops from simultaneously attempting to remove the same steady-state error in $u_{\text{dc}}^2$, which could otherwise lead to control conflicts and unintended steady-state power redistribution. Consequently, the transient protection loop provides fast transient support to limit \ac{DC-link} excursions, while the restoration loop guarantees long-term power balance and recovery, yielding a coordinated response without steady-state interference on the \ac{AC} side.

By eliminating high-frequency switching and detailed \ac{EMT} transients, the \ac{GA} design process remains computationally efficient. The resulting control set is subsequently assessed in detailed \ac{EMT} simulations in Section~\ref{sec.4_detailed_model_test}, confirming that the proposed \ac{DC-link} supervisory action remains compatible with realistic \ac{AC}-side \ac{GFM} behavior while effectively controlling the \ac{DC-link} voltage during severe transients.

\section{Control Algorithm Design Methodology}\label{sec.3_methodology}

This section covers the control design architecture and the formulation of fitness functions for obtaining an effective set of controller parameters using a \ac{GA}. It then provides qualitative remarks on the selection and robustness of the \ac{GA} hyperparameters, as well as the supervisory nature of the proposed \ac{DC-link} control. Finally, the results are presented, and the logic for acting on both the \ac{PES} and the converter output power is illustrated.

\subsection{Design with a Genetic Algorithm}\label{subsec.3_1_ga}

The parameters of controllers {\em PI-PES} and {\em PD-out} are shown in Figure~\ref{fig_block_diagram_1}. The design has been carried out with the \ac{GA} implemented in MATLAB, initially introduced in~\cite{holland1962outline} and explained in~\cite{GeneticAlgorithm}. \acp{GA} are a reliable and effective way to tune controller parameters, as shown in~\cite{joseph2022metaheuristic}. The \ac{GA} is a method for solving constrained or unconstrained optimization problems by simulating the natural selection process of biological evolution. It iteratively evolves a population of individual solutions. In each step, the algorithm randomly selects individuals from the current population and uses them as parents to produce offspring for the next generation. Depending on the values given to its parameters, a suitable solution is obtained through the evolution of successive generations.

\acp{GA} have been widely applied in power and energy systems to optimize controller parameters and improve dynamic performance. For instance, \mymarginnote{R2}\mycolor{blue}the authors in\mycolor{black}~\cite{dogruer2022design} employ a \ac{GA} to simultaneously tune the \ac{PID} gains and the fuzzy-logic scaling factors of a Fuzzy–\ac{PID} controller for an \ac{AVR} system, using a multi-objective function based on \mymarginnote{R2}\mycolor{blue}\ac{ITAE} \mycolor{black}and peak output voltage. In~\cite{hassan2023improved}, an enhanced \ac{MPPT} scheme integrates a \ac{GA} into the \ac{FOCV} method to optimally compute the $K_v$ constant under varying irradiance, thereby maximizing \ac{PV} energy extraction. Likewise, \mycolor{blue}the study in\mycolor{black}~\cite{bhukya2021parameter} applies a \ac{GA} to tune the parameters of both a \ac{PSS} and a \ac{STATCOM} controller, achieving improved damping and enhanced small-signal stability in a multi-machine power system.

A flow diagram of the methodology used, including the \ac{GA}, is depicted in Figure~\ref{fig_ga_flow_diagram}. It can be described as follows:

\begin{figure}
    \centering
    \includegraphics[width = 0.8\linewidth]{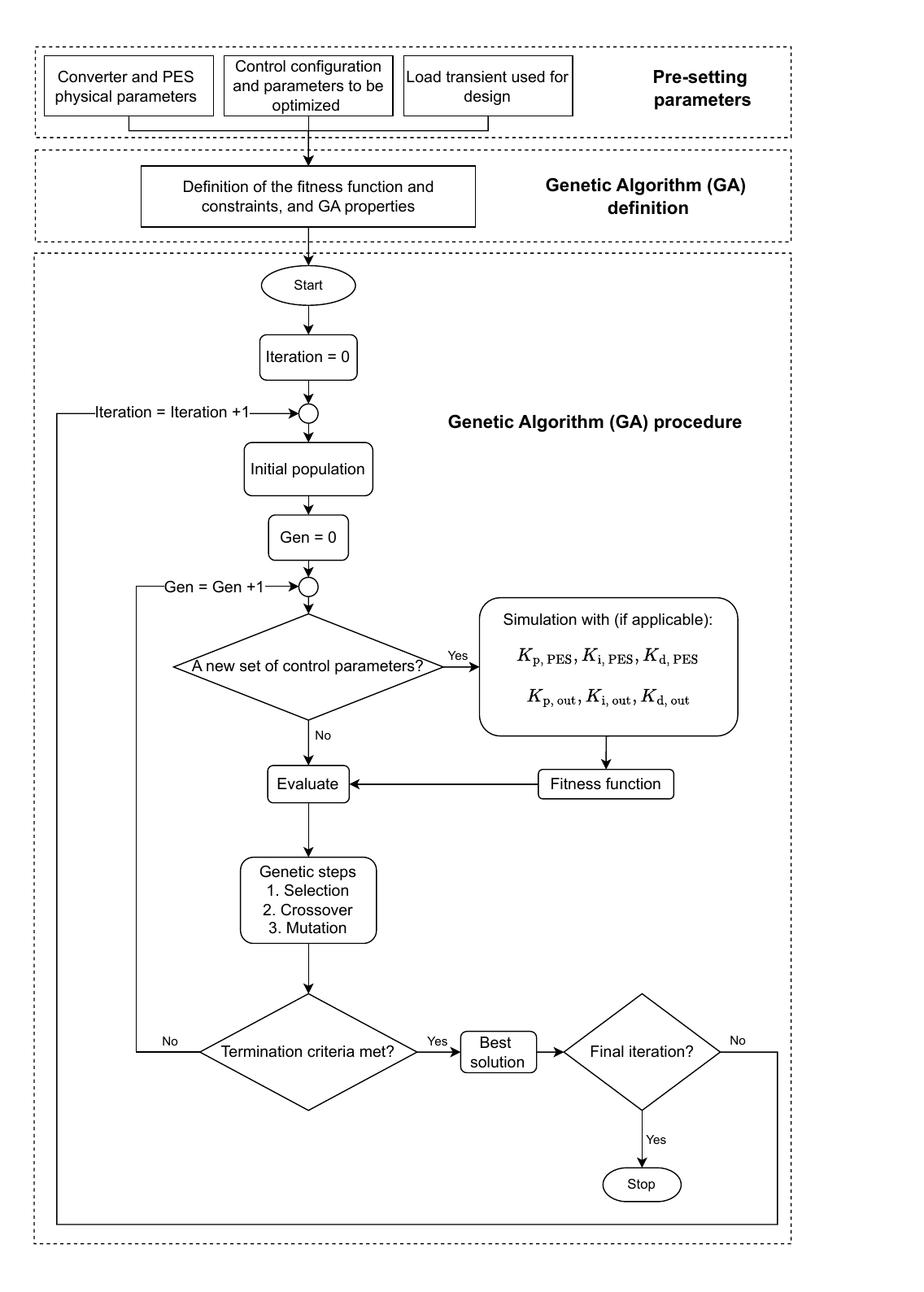}
    \caption{Flow diagram of the proposed \ac{GA}-based control design methodology}
    \label{fig_ga_flow_diagram}
    
\end{figure}

\begin{enumerate}
    \item Depending on the physical characteristics of the converter, \ac{PES} and the variables to be controlled, the following information must be provided:
    
    \begin{itemize}
        \item The pre-fixed physical and grid load parameters:
        \begin{itemize}
            \item The output power response time of the converter $T_{c}$, in seconds.
            \item The capacitance of the \ac{DC-link}, $C$, in seconds, following Equation~\eqref{eq.DC_inertia_constant}. 
            \item The feed-forward gain $K_{ff}$: set to 1 if a measurement device is available, and 0 otherwise.
            \item The \ac{PES} first-order time constant, in seconds ($T_{PES}$).
            \item The grid load setpoint that the power converter must supply, in p.u. ($p_{\text{load}}$).
        \end{itemize}
        \item The controller parameters to be optimized by the \ac{GA}, in p.u., \mymarginnote{R2}are $K_\text{p,\,PES}$ and $K_\text{i,\,PES}$ for {\em PI-PES}, and $K_\text{p,\,out}$ and $K_\text{d,\,out}$ for {\em PD-out}. $K_\text{d,\,PES}$ and $K_\text{i,\,out}$ are intentionally set to zero. However, they could be included as controller parameters in the optimization, as shown in Section~\ref{subsec.3_4_2_both_sides}, depending on the desired control behavior.
        \item The fitness function (i.e., the optimization objective) and constraints of the optimization algorithm (see Section~\ref{subsec.3_2_fitness_explanation}). 
    \end{itemize}
    
    \item After defining the pre-settings, the \ac{GA} properties must be defined to start the iterative optimization process.
    \item Finally, the \ac{GA} procedure is executed to find the most suitable controller parameters. A detailed description of the \ac{GA} steps can be found in~\cite{joseph2022metaheuristic,MathWorks_GA_Works,alhijawi2024genetic}. Nonetheless, Section~\ref{subsec.3_3_qualitative_considerations} offers further qualitative insights into the \ac{GA} characteristics and the operational principles implemented in this article.
\end{enumerate}

Selecting $T_{PES}$ is a key step in this process. While \mymarginnote{R2}\mycolor{blue}the studies in\mycolor{black}~\cite{pawar2021grid} and~\cite{guan2021scheduled} indicate that, for technologies such as \ac{PV} systems and \acp{BESS}, the \ac{DC-link} dynamics may allow energy delivery within approximately 1--10~ms, \mycolor{blue}the work in\mycolor{black}~\cite{zhou2023analysis} notes that the practical response time often lies at the upper end of this range or even exceeds it. Moreover, it is not evident that such fast power variations are always desirable. For instance, \mycolor{blue}the authors in\mycolor{black}~\cite{zhang2024coordinated} emphasize that \acp{WTG} require limits on the rate of power change to prevent mechanical and electrical stress, even when advanced control strategies are employed to provide \ac{FFR}, as in~\cite{lyu2023grid}. For these reasons, and as summarized in Table~\ref{tab_paramvalues}, we adopt a conservative value of approximately 1~s for $T_{PES}$.

In addition, it should be noted that a step change in system load, as shown in Figure~\ref{fig_block_diagram_1}, was selected as the input signal, since any mismatch between generation and consumption in the grid is perceived by the converter as a load variation. Reference~\cite{gao2025iterative} proposes state-space disturbance equations for frequency stability analysis, clarifying how load step disturbances can be equivalently represented as step disturbances at the terminals of synchronous generators and renewable units.

Finally, although a heuristic control design approach such as the \ac{GA} has been employed in this study, the proposed framework is not strictly limited to this method. Alternative optimization techniques, ranging from heuristics such as \ac{PSO} to gradient-based and grid search algorithms, could also be applied to derive the control parameters, provided the optimization problem is appropriately formulated.

\subsection{Optimization Problem: Fitness Function}\label{subsec.3_2_fitness_explanation}

Controller parameters were calculated by solving the following optimization problem: 

\noindent \hrulefill
\begin{quotation}
\begin{equation}
\underset{p, u^2} {\text{min}} \quad J= w_1 \cdot \text{IAE}_p + w_2 \cdot \text{IAE}_{u^2} + \text{penalty}
\label{eq.objfunction}
\end{equation}

subject to: 

\noindent Starting from $\text{penalty} = 0$, for $\text{signal}_i(t) \in \{ p_{\text{out}},\; p_{\text{pes-sp}},\; p_{\text{pes}} \}$:

\vspace{-1cm}

\begin{eqnarray}
\text{if } \text{signal}_i(t) > 1, & \null &
\text{penalty} += 1000 \cdot \sum \left( \text{signal}_i(t) - 1 \right), \label{eq.const_ubp} \\
\text{if } \text{signal}_i(t) < 0, & \null &
\text{penalty} += 1000 \cdot \sum \left( -\text{signal}_i(t) \right), \label{eq.const_lbp}
\end{eqnarray}

\noindent For $\Delta p_{\text{out-sp}}(t)$:

\vspace{-1cm}

\begin{eqnarray}
\text{if } \Delta p_{\text{out-sp}}(t) > 0, & \null &
\text{penalty} += 1000 \cdot \sum \Delta p_{\text{out-sp}}(t), \label{eq.const_ub_dpout} \\
\text{if } \Delta p_{\text{out-sp}}(t) < -1, & \null &
\text{penalty} += 1000 \cdot \sum \left( -1 - \Delta p_{\text{out-sp}}(t) \right), \label{eq.const_lb_dpout}
\end{eqnarray}

\noindent For $u_{\text{dc}}^2(t)$:

\vspace{-1cm}

\begin{eqnarray}
\text{if } u_{\text{dc}}^2(t) > (1.15)^2, & \null &
\text{penalty} += 1000 \cdot \sum \left( u_{\text{dc}}^2(t) - (1.15)^2\right), \label{eq.const_u2_upper} \\
\text{if } u_{\text{dc}}^2(t) < (0.7)^2,&\null &
\text{penalty} += 1000 \cdot \sum \left((0.7)^2 - u_{\text{dc}}^2(t) \right), \label{eq.const_u2_lower}
\end{eqnarray}

\noindent and, 
\begin{equation}
\begin{aligned}
& 0 \leq K_\text{p,\,PES} \leq 10,
& 0 \leq K_\text{i,\,PES} \leq 10, \\
& K_\text{d,\,PES} = 0, & 0 \leq K_\text{p,\,out} \leq 10, \\
& K_\text{i,\,out} = 0,
& 0 \leq K_\text{d,\,out} \leq 5
\end{aligned}
\label{eq.param_bounds}
\end{equation}

where $w_1 = 1.0$ and $w_2 = 1.0$, and the integral errors are defined as:
\begin{equation}
\text{IAE}_p = \int_0^T e_p(t) \, dt, \;\; e_p(t) = \left| p_{\text{out}}(t) - p_{\text{load}}(t) \right| 
\label{eq.iae_p1}
\end{equation}

\begin{equation}
\text{IAE}_{u^2} = \int_0^T e_{u^2}(t) \, dt, \;\; e_{u^2}(t) = \left| u_{\text{dc}}^2(t) - u_{\text{dc-sp}}^2(t) \right| 
\label{eq.iae_u2_1}
\end{equation}

\end{quotation}

\noindent \hrulefill

The formulation above corresponds to an optimization problem with implicit constraints. To ensure constraint satisfaction, a penalty method is employed, i.e., the fitness function includes additive terms that impose high costs when constraints are violated. This penalized fitness function is evaluated for each candidate solution representing a specific set of parameters (chromosome), and a corresponding score is assigned based on performance.

During each generation of the \ac{GA}, the algorithm evaluates the population and applies selection, crossover, and mutation operators to evolve the solutions to minimize the fitness score ($J$). The evolutionary process continues iteratively until no significant improvement is detected according to a predefined function tolerance criterion. Parameters $w_1$ and $w_2$ could be used to prioritize either $\text{IAE}_p$ or $\text{IAE}_{u^2}$, if necessary. Equation~\eqref{eq.iae_p1} calculates the \ac{IAE} between $p_{\text{out}}$ and $p_{\text{load}}$, while~\eqref{eq.iae_u2_1} calculates the \ac{IAE} between $u_{\text{dc}}^2$ and $u_{\text{dc-sp}}^2$.

Equations~\eqref{eq.const_ubp} and~\eqref{eq.const_lbp} restrict the converter output power and \ac{PES}-related signals $p_{\text{out}}$, $p_{\text{pes-sp}}$, and $p_{\text{pes}}$ to the range $[0, 1]$~p.u. Equations~\eqref{eq.const_ub_dpout} and~\eqref{eq.const_lb_dpout} restrict the incremental setpoint correction $\Delta p_{\text{out-sp}}$ provided by the {\em PD-out} controller to the range $[-1, 0]$~p.u., reflecting its role as a non-positive correction that protects the \ac{DC-link} by reducing the converter setpoint during voltage drops (see Figure~\ref{fig_block_diagram_1}). A penalty is applied if any of these limits are exceeded. An important condition is that the squared \ac{DC-link} voltage does not exceed certain limits ($0.7^2 < u_{\text{dc}}^2 < 1.15^2$) to prevent the disconnection of the power converter. Therefore, a high penalty is applied if $u_{\text{dc}}^2$ exceeds those limits as shown in~\eqref{eq.const_u2_upper} and~\eqref{eq.const_u2_lower}. Finally,~\eqref{eq.param_bounds} gives the lower and upper bounds of the controller parameters selected. $K_\text{d,\,PES}$ and $K_\text{i,\,out}$ were left out of the optimization and set to zero. When $K_\text{i,\,out} \neq 0$, the final value of $p_{\text{out}} \neq p_{\text{load}}$ when $u_{\text{dc-sp}}^2$ is driven back to its initial value, since in the steady state the converter will not respect the frequency droop imposed on the \ac{AC} side, which is not the intended response. In addition, $K_\text{d,\,PES}$ is set to zero to avoid excessive demand on the \ac{PES}.

This approach allows the \ac{GA} to explore feasible solutions; however, the qualitative remarks in Section~\ref{subsec.3_3_qualitative_considerations} are essential for a successful implementation.

\subsection{Qualitative Remarks about GA Characteristics: Hyperparameters, Optimization Search Space, Robustness and Convergence}\label{subsec.3_3_qualitative_considerations}

The \ac{GA} hyperparameters used in the optimization process are summarized in Table~\ref{tab_ga_params}. To balance the trade-off between computational burden, accuracy, and algorithmic reliability, the proposed methodology employs a heterogeneous rounding technique, mapping continuous search variables to a high-resolution discrete combinatorial domain. Specifically, a scaling vector is applied to enforce a precision of two decimal places for the proportional and integral gains ($K_\text{p,\,PES}$, $K_\text{p,\,out}$, and $K_\text{i,\,PES}$), and three decimal places for the derivative gain ($K_\text{d,\,out}$). This controlled reduction of the search space significantly accelerates the optimization process. As a result, a relatively compact population size of 50 individuals is sufficient to explore the viable space effectively, while a stringent function tolerance ($10^{-10}$) ensures high precision. To handle this discrete optimization phase, the algorithm employs the default integer mutation logic provided by MATLAB (\texttt{mutationpower}). Furthermore, a two-iteration verification process is executed, with the second iteration building on the best candidates identified at the end of the first. The maximum number of generations per iteration is set to 200, resulting in a total of 400 potential generations for each complete \ac{GA} execution.

\begin{table}[!ht]
    \caption{\ac{GA} hyperparameters applied}
    \centering
    \begin{tabular}{p{0.5\linewidth}p{0.40\linewidth}}
    \toprule
        \textbf{\ac{GA} option} & \textbf{Selection} \\ 
        \midrule
        Function tolerance (FT) & $10^{-10}$ \\ 
        Constraint tolerance (CT) & $10^{-3}$ \\
        Max. stall generation number (MSGN) & 50 \\
        Population elite count & 5\% \\
        Max. number of generations & 200 \\
        Population size & 50 \\
        Algorithm & Default integer mutation \\
        Initial population creation & Deterministic \\
        Total iterations & 2 \\
    \bottomrule
    \end{tabular}
    \label{tab_ga_params}
\end{table}

It is worth noting that the \ac{GA} is deliberately applied to a focused optimization search space. As detailed in Section~\ref{subsec.3_4_1_base_case}, only four \ac{DC-link} control parameters are tuned within constrained bounds. Following the assumptions established in Section~\ref{sec.2_problem_and_case_study}, this study does not aim to optimize the complete set of \ac{GFM-VSC} parameters, but rather to find a suitable parameter set exclusively for the \ac{DC-link} control. Furthermore, the resulting parameter set is intrinsically dependent on the specific operating point. The obtained gains rely on the active-power operating conditions, the available power reserve of the \ac{PES}, the \ac{DC-link} capacitance, and the established control bounds. Consequently, this strategy is not intended to find a globally optimal solution that is universally independent of the operating point; if these system conditions change significantly, the parameters must be re-optimized. These scenarios are further detailed and analyzed in Section~\ref{subsec.3_4_3_additional_cases}.

The \ac{GA} execution terminates when the relative improvement in the objective function remains below the predefined function tolerance over a consecutive number of stall generations, or when the absolute maximum generations limit is reached, as illustrated in Figure~\ref{fig_ga_flow_diagram}. It should be noted that while the constraint tolerance defines the solution's feasibility, it does not serve as an independent stopping criterion. Instead, if the algorithm concludes its search—due to stalling or reaching the generation limit—and the best solution found still exhibits a maximum constraint violation exceeding the constraint tolerance, the result is classified as infeasible~\cite{MathWorks_GA_Works}. This indicates that the search failed to reach a region satisfying all operational boundaries within the allocated generations.

Considering the defined search space, the operating point, and the selected hyperparameters, a solid solution is obtained upon convergence of the base case, as analyzed in Section~\ref{subsec.3_4_1_base_case}. In this scenario, the objective function error is minimized, and the variation of the control parameters during the final generations is practically negligible. The algorithm successfully converges because the improvement of the objective function falls below the imposed function tolerance for the maximum stall generation number, indicating that the optimization goals are fully met.

Conversely, if the algorithm fails to identify an optimal or feasible solution, several underlying causes must be evaluated. First, an inadequate \ac{GA} configuration could lead to infeasible or sub-optimal solutions. If possible, this can be mitigated by tuning the \ac{GA} hyperparameters (e.g., increasing the population size and maximum generations, or reducing tolerances), performing multi-start runs with different random initial populations, or executing additional sequential algorithm iterations seeded with the best solution found in the previous run. These strategies inherently increase the computational burden and assume that a feasible solution actually exists for the analyzed operating point. Second, the failure might originate from an incorrect or overly restrictive specification of the optimization problem itself. Third, the lack of convergence could reflect a physical impossibility within the system hardware or sizing; for instance, the \ac{DC-link} capacitor might be excessively small, or the \ac{PES} might be too slow to counteract a severe disturbance. Lastly, the selected control variables may be insufficient to achieve the desired dynamic performance. This last scenario is represented in Section~\ref{subsec.3_4_2_both_sides}.

\subsection{Results}\label{subsec.3_4_Results}

This section presents the results obtained using the \ac{GA}. A base case with a representative 0.1~p.u. load increase is first reported, followed by an illustration of the rationale for coordinating the \ac{PES} power and the converter output power in response to \ac{DC-link} deviations. Finally, the control performance is evaluated under additional operating points and load imbalances to demonstrate its effectiveness across different operating conditions.

\subsubsection{Base Case}\label{subsec.3_4_1_base_case}

This section reports the response of the \ac{GA}-tuned control to a 0.1~p.u. increase in \ac{AC}-side active-power demand. The values of the predefined parameters and the optimized controller gains are shown in Table~\ref{tab_paramvalues}. Table~\ref{tab_ga_params} shows the \ac{GA} parameters and options selected, besides the default ones from MATLAB. $T_{PES}$ and $T_{c}$ are conservatively slow compared to typical values reported in the literature (e.g., in reference~\cite{girona2024resource}). In addition, the equivalent \ac{DC-link} capacitance is approximately 1.3~mF (derived from a time constant of 0.02~s) for a power converter rated at 15~kVA and 680~V DC, similar to the converter used in~\cite{gothner2021harmonic}. This value is comparable to the 1~mF capacitor employed on the \ac{DC} side in that reference. The conversion between seconds and Farads follows Equation~\eqref{eq.DC_inertia_constant}, as explained in Section~\ref{subsubsec.2_1_4_dc_link}.

\begin{figure}
    \centering
    \includegraphics{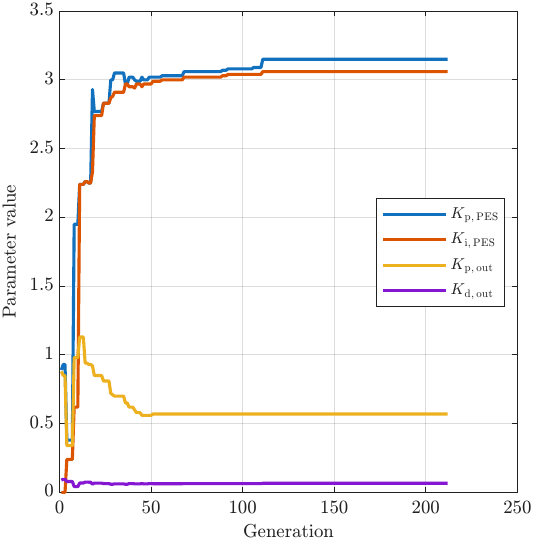}
    \vspace{-3mm}
    \caption{Evolution of controller parameters during the \ac{GA} optimization}
    \label{fig_control_parameters_evolution}
\end{figure}

Figure~\ref{fig_control_parameters_evolution} shows the evolution of the parameters involved in the optimization process as the \ac{GA} progresses. Clearly, from approximately generation 110 onward (out of 400), the controller parameters barely change. The first iteration stops at generation 160, and the second iteration stops after 51 generations, making a total of 211 generations.

Figure~\ref{fig_comb_control_response} shows the first second of a simulation run with the simplified model after a 0.1~p.u. stepwise increment in the \ac{AC} load takes place. The settling time is marked with a dashed vertical line when $u_{\text{dc}}^2$, $p_{\text{out}}$, and $p_{\text{pes}}$ reach 99\% of their final values after the power step occurs. As there is not much energy stored in the \ac{DC-link} capacitor, $u_{\text{dc}}^2$ initially decays to approximately $0.84$~p.u. (see the upper subplot). It then quickly recovers to $0.99$~p.u. in about $530$~\mymarginnote{R2}\mycolor{blue}ms \mycolor{black}after the disturbance. The fast recovery is due to {\em PD-out} (which acts on the output power of the converter). Eventually, it settles to $u_{\text{dc-sp}}^2$, owing to the {\em PI-PES} controller.

The lower subplot of Figure~\ref{fig_comb_control_response} shows $p_{\text{out}}$, $p_{\text{pes-sp}}$, and $p_{\text{pes}}$ of the converter. First, the converter attempts to deliver the power that is demanded from the grid. However, as the \ac{DC-link} voltage rapidly decays, the power converter quickly reduces its output power to $0.501$~p.u. Once the \ac{DC-link} recovers, the output power stabilizes at $0.59$~p.u., in around $486$~\mymarginnote{R2}\mycolor{blue}ms\mycolor{black}. Simultaneously, the \ac{PES} power setpoint, $p_{\text{pes-sp}}$, increases sharply following the disturbance until reaching its upper saturation limit ($\approx 1$~p.u.). Subsequently, it gradually decays to align with the new power demand required by the grid. In contrast, the actual power delivered by the \ac{PES}, $p_{\text{pes}}$, is governed by its inherent slow dynamics (defined by $T_{PES}$); consequently, it ramps up progressively until it eventually matches the converter output power, $p_{\text{out}}$.

The tuning of the \ac{GA} allows the \ac{DC-link} control to operate effectively, maintaining $u_{\text{dc}}^2$ within safe operational limits while delivering the required output power ($p_{\text{out}}$) according to grid demand in steady state.

\subsubsection{Logic Behind Acting on Both the PES and the Output of the Power Converter}\label{subsec.3_4_2_both_sides}

\begin{figure}
    \centering
    \includegraphics[]{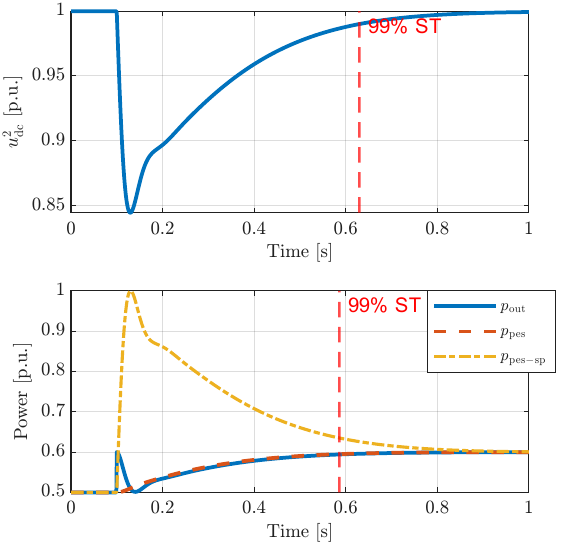}
    \vspace{-3mm}
    \caption{Control response of $u_{\text{dc}}^2$, $p_{\text{out}}$, $p_{\text{pes}}$ and $p_{\text{pes-sp}}$}
    \label{fig_comb_control_response}
    
\end{figure}

\begin{figure}
    \centering
   \includegraphics[]{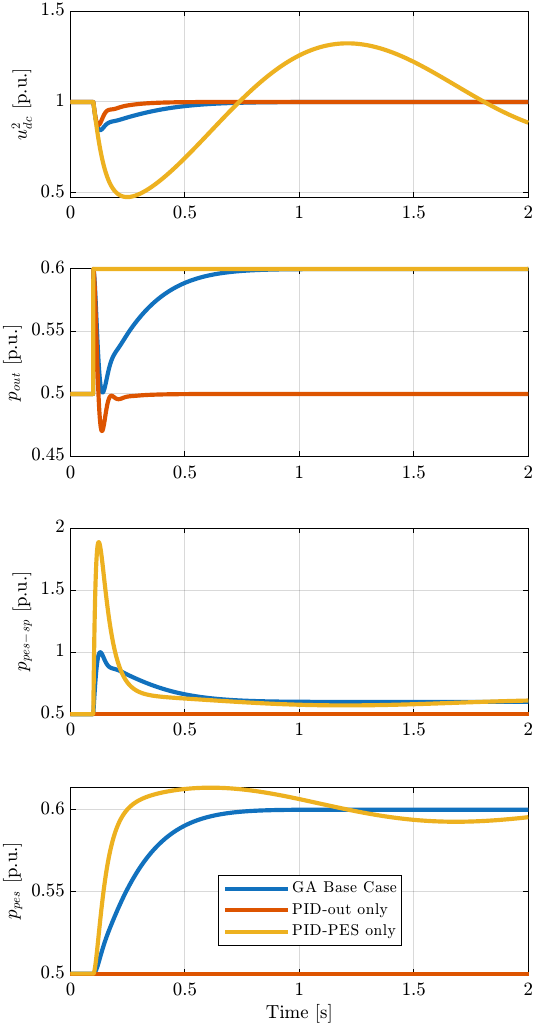}
    \caption{Comparison of control response of GA Base Case, PID-out only, and PID-PES only}
    \label{fig_comparison_zooms}
\end{figure}

The rationale for coordinating the \ac{PES} power and the converter output power is presented for the same disturbance considered in Section~\ref{subsec.3_4_1_base_case}. The benefits of acting on both the \ac{PES} and the converter output power can be observed in Figure~\ref{fig_comparison_zooms}, where three parameter configurations are compared: (a) the \ac{GA} Base Case from Figure~\ref{fig_comb_control_response}; (b) {\em PID-out}, featuring only the converter output power loop as a \ac{PID} controller (leaving $K_\text{p,i,d,\,PES} = 0$); and (c) {\em PID-PES}, featuring only the \ac{PES} power loop as a \ac{PID} controller (setting $K_\text{p,i,d,\,out} = 0$). Note that for cases (b) and (c), all three gains (proportional, integral, and derivative) of the analyzed \ac{PID} loops are optimized through the \ac{GA}. In contrast, for case (a), $K_\text{d,\,PES}$ and $K_\text{i,\,out}$ are set to zero. The set of parameters obtained for each case is shown in Table~\ref{tab_paramvalues_pid_configs}.

In case (b), where {\em PID-out} controller parameters are optimized, the restoration of $u_{\text{dc}}^2$ after the power disturbance is fast ($\approx 211$~ms). The integral component of the {\em PID-out} controller ensures that the reference setpoint $u_{\text{dc-sp}}^2$ is tracked. However, the resulting power response is unsatisfactory because the converter output power does not follow the setpoint requested from the grid. Since the \ac{PES} remains inactive, the converter output power is unchanged in the steady state, leaving the inertial response constrained by the limited energy stored in the \ac{DC-link} capacitor, without \ac{PFR} capability. As a result, the converter output power fails to track the grid power step in the steady state.

On the other hand, in case (c), where the {\em PID-PES} parameters are tuned, $p_{\text{out}}$ is delivered instantaneously to the grid. However, two factors demonstrate the infeasibility of this solution. First, the required power setpoint for the \ac{PES} ($p_{\text{pes-sp}}$) is infeasible because it exceeds the $1$~p.u. upper bound. Furthermore, the actual power delivered by the \ac{PES} ($p_{\text{pes}}$) first reaches the grid demand within approximately $150$~ms. This response speed deviates from the expected physical dynamics governed by the time constant $T_{PES}$, due to the unrealistic and excessive setpoint to the \ac{PES}. Second, the restoration of $u_{\text{dc}}^2$ is excessively slow because the {\em PID-PES} controller is not sufficiently fast, and it also causes the \ac{DC-link} voltage to drop below the lower limit and almost reach the upper bound.

In both cases (b) and (c), the \ac{GA} reached the predefined maximum number of stall generations, terminating with a relatively high objective function value, failing to find a solution that complies with the restrictions. These results demonstrate that \ac{DC-link} control must act on both the \ac{PES} and the converter output power to satisfy operational constraints and maintain the power converter's grid connectivity.

\begin{table}[!ht]
    \caption{Parameters applied on the base case} 
    \centering 
    \begin{tabular}{c|c|c} \toprule 
        \multicolumn{3}{c}{\textbf{System Parameters}} \\
        \midrule 
         $C$ (s) = 0.02 & $u_\text{dc-sp}^2$ = 1 & $p_{\text{out-sp}}$ (p.u.) = 0.5 \\
         $p_\text{load-ini}$ (p.u.) = 0.5 & $p_\text{load-fin}$ (p.u.) = 0.6 & $T_{PES}$ (s) = 1 \\ 
         $T_c$ (s) = 0.1 & $K_\text{ff}$ = 0 & \\ 
        \midrule
        \multicolumn{3}{c}{\textbf{DC-link control design}} \\
        \midrule
         $K_\text{p,\,PES}$ (p.u.) = 3.15 & 
         $K_\text{i,\,PES}$ (p.u.) = 3.06 & 
         $K_\text{d,\,PES}$ (p.u.) = 0 \\
         $K_\text{p,\,out}$ (p.u.) = 0.57 & 
         $K_\text{i,\,out}$ (p.u.) = 0 & 
         $K_\text{d,\,out}$ (p.u.) = 0.066 \\ 
    \bottomrule 
    \end{tabular} 
    \label{tab_paramvalues}
\end{table}

\begin{table}[!ht]
    \caption{\ac{DC-link} control designs used in the comparison}
    \centering
    \begin{tabular}{c|c|c|c}
    \toprule
    {\bf Condition} & {\bf Case (a)} & {\bf Case (b)} &{\bf Case (c)} \\
    \midrule
        
        $K_\text{p,\,PES}$ & 3.15  & 0     &  0.26 \\
        $K_\text{i,\,PES}$ & 3.06  & 0     &  1.53 \\
        $K_\text{d,\,PES}$ & 0     & 0     &  0.171 \\
        $K_\text{p,\,out}$ & 0.57  & 1.90  &  0 \\
        $K_\text{i,\,out}$ & 0     & 10.00 &  0 \\
        $K_\text{d,\,out}$ & 0.066 & 0.077 &  0 \\
    \bottomrule
    \end{tabular}
    \label{tab_paramvalues_pid_configs}
\end{table}

\subsubsection{Cases from Different Operating Points and Power Imbalance Magnitudes}\label{subsec.3_4_3_additional_cases}

This section presents the implementation of the \ac{GA}, following the same procedure as in Section~\ref{subsec.3_4_1_base_case}, under different operating conditions with different load steps, such as $p_\text{load-ini} = [0.3, 0.5, 0.7]$~p.u. and $\Delta p_\text{load} = [0.05, 0.1, 0.2]$~p.u., where $p_\text{load-fin}=p_\text{load-ini}+\Delta p_\text{load}$ is the final load power requested by the grid in the steady state. A total of nine cases were implemented by assigning a specific set of \ac{DC-link} parameters ($K_\text{p,\,PES}$, $K_\text{i,\,PES}$, $K_\text{p,\,out}$, $K_\text{d,\,out}$) to each operating point and disturbance. All other parameters remain consistent with the values provided in Tables~\ref{tab_ga_params} and~\ref{tab_paramvalues}.

As a result of optimizing the \ac{GA} through the nine cases, Tables~\ref{tab_min_udc2} and~\ref{tab_settle_pout} are obtained. Table~\ref{tab_min_udc2} presents the minimum value of $u_\text{dc}^2$ when simulating $\Delta p_\text{load}$ at each initial operating point. Considering that $u_\text{dc}^2$ must be greater than or equal to $0.7^2$~p.u., this shows that the optimized control parameters are robust in each case. In addition, Table~\ref{tab_settle_pout} reports the \ac{ST}, defined as the time at which the converter output power reaches $99\%$ of its steady-state value, for each case. 

\begin{table}[!ht]
    \caption{Minimum value of $u_\text{dc}^2$ (p.u.)}
    \centering
    \begin{tabular}{c|c|c|c}
    \toprule
    {\bf $u_\text{dc-min}^2$} & {\bf $\Delta p_\text{load} = 0.05$} & {\bf $\Delta p_\text{load} = 0.10$} & {\bf $\Delta p_\text{load} = 0.20$} \\
    \midrule
        $p_\text{load-ini} = 0.30$ (p.u.) & $0.9015$ & $0.7618$ & $0.7090$ \\
        $p_\text{load-ini} = 0.50$ (p.u.) & $0.8988$ & $0.8445$ & $0.6834$ \\
        $p_\text{load-ini} = 0.70$ (p.u.) & $0.9126$ & $0.8551$ & $0.6971$ \\
    \bottomrule
    \end{tabular}
    \label{tab_min_udc2}
\end{table}

\begin{table}[!ht]
    \caption{99\% settling time of $p_\text{out}$ (s)}
    \centering
    \begin{tabular}{c|c|c|c}
    \toprule
    {\bf $\text{ST}_{99\%}$ (s)} & {\bf $\Delta p_\text{load} = 0.05$} & {\bf $\Delta p_\text{load} = 0.10$} & {\bf $\Delta p_\text{load} = 0.20$} \\
    \midrule
        $p_\text{load-ini} = 0.30$ (p.u.) & $0.351$ & $0.385$ & $0.800$ \\
        $p_\text{load-ini} = 0.50$ (p.u.) & $0.299$ & $0.486$ & $0.834$ \\
        $p_\text{load-ini} = 0.70$ (p.u.) & $0.300$ & $0.643$ & $1.580$ \\
    \bottomrule
    \end{tabular}
    \label{tab_settle_pout}
\end{table}

The results obtained validate that \ac{GA} optimization yields consistent control parameters that ensure stable performance across diverse operating points and disturbance scenarios. More importantly, the analysis reveals that the converter's grid-support capability is fundamentally determined by the synergy between the \ac{PES} and the \ac{DC-link} energy reserves. In this regard, a greater power imbalance, coupled with lower available reserves, results in a slower \ac{AC}-side response, while larger imbalances further exacerbate the \ac{DC}-side voltage drop. Specifically, the energy extraction dynamics, strictly governed by the \ac{PES} time constant ($T_{PES}$), determine the converter's effectiveness in mitigating such grid disturbances.

Ultimately, the practical selection of control parameters, as established in Section~\ref{subsec.3_3_qualitative_considerations}, is a delicate task that depends on several factors: the operating point (available reserves), the expected disturbance magnitude, the size of the \ac{DC-link} capacitor, and the dynamics of the \ac{PES}. While the pre-disturbance conditions are known, the scale of an upcoming power imbalance is inherently unpredictable. Consequently, a conservative design approach could be adopted by tuning the controller to withstand the largest credible power disturbance. This strategy prioritizes maintaining converter connectivity and preventing tripping during severe events, even if it results in a slightly slower \ac{AC}-side power response, provided that sufficient energy reserves are available.

\section{Detailed Full-Order Model Validation}\label{sec.4_detailed_model_test}

This section validates the proposed \ac{DC-link} control using an \ac{EMT} model of the \ac{GFM-VSC}, which includes a typical \ac{VSM} representation (with virtual inertia and damping including a washout filter) and current and voltage control loops. The control parameters are tuned so that the converter follows changes in the active power setpoint with a first-order, critically damped response, as explained in Section~\ref{subsubsec.2_1_2_converter_dynamics}. The full details on the model and parameters used are included in~\cite{VFlexPGitHubCase}. This complete model, or detailed full-order model, is implemented using the tool described in~\cite{tomas2025vector} to replicate the study presented in Section~\ref{subsec.3_4_1_base_case}, and the results of the detailed simulation and the simplified model are compared in this section. The standard \ac{GFM-VSC} model (without the \ac{DC-link} control) included in the tool described in~\cite{tomas2025vector} was previously validated against laboratory measurements obtained at \ac{IMDEA} Energy in~\cite{tomas2026improvements}.

\begin{figure}
    \centering
    \includegraphics[width = 0.8\linewidth]{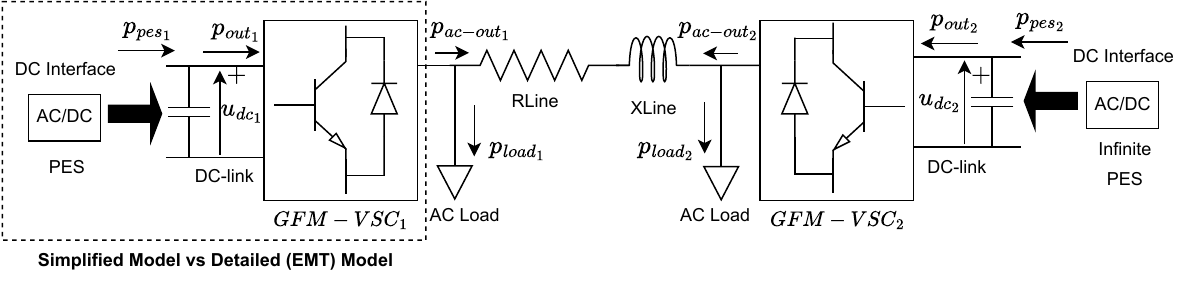}
    \caption{Case study of the detailed model system: two \acp{GFM-VSC} interconnected}
    \label{fig_simp_vs_detailed_case}
\end{figure}

\begin{figure}
    \centering
    \includegraphics[width=0.6\textwidth]{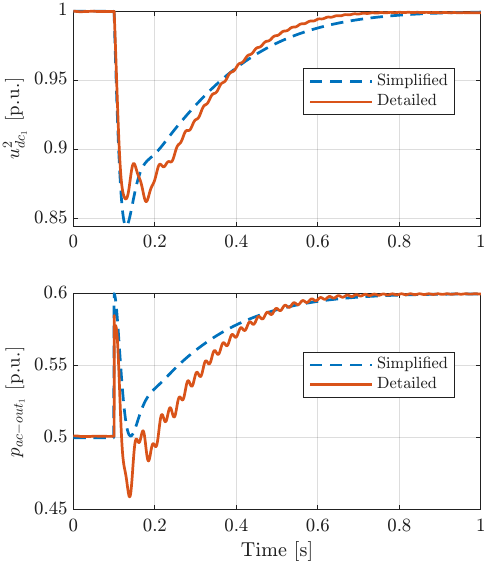}
    \caption{Control response comparison between models}
    \label{fig_simp_vs_detailed_comparison}
    
\end{figure}

Figure~\ref{fig_simp_vs_detailed_case} depicts the case system for the validation of the results in a detailed model. It is composed of two \acp{GFM-VSC} connected through a line, and a load connected at each line end. The following assumptions are considered:
\begin{itemize}
    \item $\ac{GFM-VSC}_{1}$ includes a \ac{PES}, the \ac{DC-link} capacitor dynamics, and the \ac{DC-link} control presented in Section~\ref{sec.3_methodology}. The values of the parameters of the \ac{DC-link} control are those indicated in Table~\ref{tab_paramvalues}. Both the simplified model from Section~\ref{sec.3_methodology} and the detailed full-order \ac{EMT} model of this $\ac{GFM-VSC}_{1}$ are simulated and compared in Figure~\ref{fig_simp_vs_detailed_comparison} to validate the proposed \ac{DC-link} control.
    \item $\ac{GFM-VSC}_{2}$ has a \ac{PES} with infinite and instantaneous power provision, and no \ac{DC} dynamics.
    \item Each \ac{GFM-VSC} is initially delivering 0.5~p.u. Each load consumes 0.5~p.u.
    \item A 0.1~p.u. step increase in \ac{AC} Load~1 (applied to both the simplified and the \ac{EMT} models) at $t = 0.1$~s is simulated.
\end{itemize}

The first and second subplots in Figure~\ref{fig_simp_vs_detailed_comparison} show the $u_\text{dc}^2$ and $p_{\text{out}}$ responses of the $\ac{GFM-VSC}_{1}$ in both the simplified and detailed models. As shown in both subplots, the results obtained with the simplified and the detailed models match well, considering the higher number of state variables in the detailed model. This demonstrates that the model proposed to design the \ac{DC-link} control is adequate, and it confirms the validity of the results.

\section{Conclusions and Future Work}\label{sec.5_conclusions}

This article proposes a \ac{DC-link} voltage supervisory control strategy for \acp{GFM-VSC} that coordinates both the converter output power setpoint and, when available, the \ac{PES} reserve power setpoint, together with a methodology to tune the control parameters. The control is optimized to prevent converter disconnection by keeping the \ac{DC-link} voltage within admissible limits while coordinating with the \ac{GFM-VSC} that provides inertial and primary-frequency support. The optimization objective explicitly accounts for both \ac{DC-link} voltage excursions and converter output-power tracking. The proposed \ac{DC-link} control layer is first designed using a simplified model and then validated in a detailed \ac{EMT} model. Despite the simplified model adopted to make the design process more tractable, the detailed \ac{EMT} simulation results closely match the response predicted by the simplified model, confirming the adequacy of the proposed reduction. The results show that the \ac{GFM-VSC}'s achievable output power response is strongly influenced by the mismatch between \ac{PES} power provision and the available \ac{DC-link} capacitance. Future research will pursue several key directions to extend the findings of this study. One research line focuses on laboratory-based experimental validation to corroborate the proposed \ac{DC-link} control performance. Another key area involves scaling the analysis to investigate collective dynamics and multi-machine interactions in larger power systems. Efforts will also be directed toward extending the design methodology into a globally robust framework, ensuring performance under severe disturbances such as short-circuit faults and varying short-circuit ratios. Lastly, future work will aim to establish a rigorous equivalence between the virtual inertia of \acp{GFM-VSC} and the dynamics of traditional \acp{SG}.

\section*{Acknowledgments}
This work is part of the Project PID2021-125628OB-C21 funded by
MICIU\slash AEI\slash 10.13039\slash 501100011033.

\section*{Declaration of Generative AI Use in the Writing Process}
During the preparation of this work, the authors used a large language model, specifically Claude (developed by Anthropic), for proofreading and language polishing. After using this tool, the authors reviewed and edited the content as needed and take full responsibility for the publication's content.

\bibliographystyle{IEEEtran}
\begingroup
\footnotesize
\setlength{\parskip}{0.5pt}
\begin{spacing}{1}
\bibliography{PES_GFM}

@misc{VFlexPGitHubCase,
  title     = {{Grid-Forming Voltage-Source Converter vFlexP Case with PES and DC-link Control}},
  author    = {De Paolis Robles, Carlo and Tom{\'a}s-Mart{\'\i}n, Andr{\'e}s},
  note      = {{GitHub} repository},
  year      = {2026},
  publisher = {GitHub},
  howpublished = {\url{https://github.com/carlodpr/GFM-VSC-DC-link-control---PES.git}}
}

@misc{MathWorks_GA_Works,
  author = {{The MathWorks, Inc.}},
  title  = {How the {G}enetic {A}lgorithm {W}orks},
  year   = {2026},
  url    = {https://es.mathworks.com/help/gads/how-the-genetic-algorithm-works.html},
  note   = {Accessed: May 5, 2026}
}

@article{avila2026impact,
  title={Impact on transient stability of self-synchronisation control strategies in grid-forming power converters},
  author={{\'A}vila-Mart{\'\i}nez, R{\'e}gulo E and Guillaud, Xavier and Renedo, Javier and Rouco, Luis and Garcia-Cerrada, Aurelio and Sigrist, Lukas},
  journal={International Journal of Electrical Power \& Energy Systems},
  volume={174},
  pages={111540},
  year={2026},
  publisher={Elsevier}
}

@article{shafique2025dc,
  title={DC voltage control with grid-forming capability for enhancing stability of HVDC system},
  author={Shafique, Ghazala and Boukhenfouf, Johan and Gruson, Fran{\c{c}}ois and Colas, Fr{\'e}d{\'e}ric and Guillaud, Xavier},
  journal={Journal of Modern Power Systems and Clean Energy},
  volume={13},
  number={1},
  pages={66--78},
  year={2025},
  publisher={SGEPRI}
}

@article{klaes2025immunity,
  title={Immunity of grid-forming control without energy storage to transient changes of grid frequency and phase},
  author={Klaes, Norbert R and Fortmann, Jens},
  journal={IEEE Open Journal of the Industrial Electronics Society},
  volume={6},
  pages={265--276},
  year={2025},
  publisher={IEEE}
}

@article{tomas2026improvements,
  title={Improvements of a multi-agent secondary controller for reconnecting a microgrid to the main grid},
  author={Tom{\'a}s-Mart{\'\i}n, Andr{\'e}s and Rold{\'a}n-P{\'e}rez, Javier and Jankovic, Njegos and Yag{\"u}e, Sauro and Sigrist, Lukas and Garc{\'\i}a-Cerrada, Aurelio},
  journal={International Journal of Electrical Power \& Energy Systems},
  volume={177},
  pages={111797},
  year={2026},
  publisher={Elsevier}
}

@incollection{rouco2013selective,
  title={Selective modal analysis},
  author={Rouco, Luis and Pagola, FL and Verghese, George C and P{\'e}rez-Arriaga, Ignacio J},
  booktitle={Power system coherency and model reduction},
  pages={199--258},
  year={2013},
  publisher={Springer}
}

@misc{entsoe2025_phaseII_news,
  author       = {{ENTSO-E}},
  title        = {ENTSO-E publishes Phase II Technical Report on Grid Forming Requirements},
  year         = {2025},
  month        = nov,
  day          = {4},
  howpublished = {ENTSO-E News},
  url          = {https://www.entsoe.eu/news/2025/11/04/entso-e-publishes-phase-ii-technical-report-on-grid-forming-requirements/},
  note         = {Last accessed: 2025-12-18}
}

@article{tayyebi2020frequency,
  title={Frequency stability of synchronous machines and grid-forming power converters},
  author={Tayyebi, Ali and Gro{\ss}, Dominic and Anta, Adolfo and Kupzog, Friederich and D{\"o}rfler, Florian},
  journal={IEEE Journal of Emerging and Selected Topics in Power Electronics},
  volume={8},
  number={2},
  pages={1004--1018},
  year={2020},
  publisher={IEEE}
}

@article{zhang2024coordinated,
  title={Coordinated frequency modulation control strategy of wind power and energy storage considering mechanical load optimization},
  author={Zhang, Chaoyu and Li, Jiabin and Liu, Shiyi and Hu, Peng and Feng, Jiangzhe and Ren, Haoyang and Zhang, Ruizhe and Jia, Jiaoxin},
  journal={Energies},
  volume={17},
  number={13},
  pages={3198},
  year={2024},
  publisher={MDPI}
}

@article{dogruer2022design,
  title={Design and robustness analysis of fuzzy PID controller for automatic voltage regulator system using genetic algorithm},
  author={Dogruer, Tufan and Can, Mehmet Serhat},
  journal={Transactions of the Institute of Measurement and Control},
  volume={44},
  number={9},
  pages={1862--1873},
  year={2022},
  publisher={SAGE Publications Sage UK: London, England}
}

@article{bhukya2021parameter,
  title={Parameter tuning of PSS and STATCOM controllers using genetic algorithm for improvement of small-signal and transient stability of power systems with wind power},
  author={Bhukya, Jawaharlal and Mahajan, Vasundhara},
  journal={International Transactions on Electrical Energy Systems},
  volume={31},
  number={7},
  pages={e12912},
  year={2021},
  publisher={Wiley Online Library}
}

@article{hassan2023improved,
  title={An improved genetic algorithm based fractional open circuit voltage MPPT for solar PV systems},
  author={Hassan, Aakash and Bass, Octavian and Masoum, Mohammad AS},
  journal={Energy Reports},
  volume={9},
  pages={1535--1548},
  year={2023},
  publisher={Elsevier}
}

@article{joseph2022metaheuristic,
  title={Metaheuristic algorithms for PID controller parameters tuning: Review, approaches and open problems},
  author={Joseph, Stephen Bassi and Dada, Emmanuel Gbenga and Abidemi, Afeez and Oyewola, David Opeoluwa and Khammas, Ban Mohammed},
  journal={Heliyon},
  volume={8},
  number={5},
  year={2022},
  publisher={Elsevier}
}

@article{alhijawi2024genetic,
  title={Genetic algorithms: Theory, genetic operators, solutions, and applications},
  author={Alhijawi, Bushra and Awajan, Arafat},
  journal={Evolutionary Intelligence},
  volume={17},
  number={3},
  pages={1245--1256},
  year={2024},
  publisher={Springer}
}

@article{zhang2021grid,
  title={Grid forming converters in renewable energy sources dominated power grid: Control strategy, stability, application, and challenges},
  author={Zhang, Haobo and Xiang, Wang and Lin, Weixing and Wen, Jinyu},
  journal={Journal of modern power systems and clean energy},
  volume={9},
  number={6},
  pages={1239--1256},
  year={2021},
  publisher={SGEPRI}
}

@article{luo2023design,
  title={Design-oriented analysis of DC-link voltage control for transient stability of grid-forming inverters},
  author={Luo, Cheng and Liu, Teng and Wang, Xiongfei and Ma, Xikui},
  journal={IEEE Transactions on Industrial Electronics},
  volume={71},
  number={4},
  pages={3698--3707},
  year={2023},
  publisher={IEEE}
}

@article{shen2023transient,
  title={Transient stability analysis and design of VSGs with different DC-link voltage controllers},
  author={Shen, Chao and Gu, Wei and Sheng, Wanxing and Liu, Keyan},
  journal={CSEE Journal of Power and Energy Systems},
  volume={10},
  number={2},
  pages={593--604},
  year={2023},
  publisher={CSEE}
}

@article{tian2023two,
  title={Two-stage PV grid-connected control strategy based on adaptive virtual inertia and damping control for DC-link capacitor dynamics self-synchronization},
  author={Tian, Aina and Wu, Yang and Hu, Zhaorui and Wang, Zhikun and Wu, Tiezhou and Jiang, Jiuchun and Peng, Zinan},
  journal={Journal of Energy Storage},
  volume={72},
  pages={108659},
  year={2023},
  publisher={Elsevier}
}

@article{qin2024novel,
  title={A novel DC-link voltage synchronous control with enhanced inertial capability for full-scale power conversion wind turbine generators},
  author={Qin, Yao and Wang, Han and Zhou, Dangsheng and Deng, Zhenyan and Zhang, Jianwen and Cai, Xu},
  journal={IET Renewable Power Generation},
  volume={18},
  number={4},
  pages={690--705},
  year={2024},
  publisher={Wiley Online Library}
}

@article{kryonidis2023use,
  title={Use of ultracapacitor for provision of inertial response in virtual synchronous generator: Design and experimental validation},
  author={Kryonidis, Georgios C and Mauricio, Juan Manuel and Malamaki, Kyriaki-Nefeli D and Barrag{\'a}n-Villarejo, Manuel and de Paula Garc{\'\i}a-L{\'o}pez, Francisco and Matas-Diaz, Francisco Jesus and Maza-Ortega, Jose Maria and Demoulias, Charis S},
  journal={Electric Power Systems Research},
  volume={223},
  pages={109607},
  year={2023},
  publisher={Elsevier}
}

@article{gross2022energy,
  title={Energy management in converter-interfaced renewable energy sources through ultracapacitors for provision of ancillary services},
  author={Gross, Andrei Mihai and Malamaki, Kyriaki-Nefeli and Barrag{\'a}n-Villarejo, Manuel and Kryonidis, Georgios C and Matas-D{\'\i}az, Francisco Jes{\'u}s and Gkavanoudis, Spyros I and Mauricio, Juan Manuel and Maza-Ortega, Jos{\'e} Mar{\'\i}a and Demoulias, Charis S},
  journal={Sustainable Energy, Grids and Networks},
  volume={32},
  pages={100911},
  year={2022},
  publisher={Elsevier}
}

@article{xu2025stability,
  title={Stability analysis and control design of grid-forming converters with dc-link effect},
  author={Xu, Chenhang and Zou, Zhixiang and Liu, Xinlei and Huang, Meng and Chen, Wu and Wang, Zheng},
  journal={IEEE Transactions on Power Electronics},
  year={2025},
  publisher={IEEE}
}

@article{karunaratne2023nonlinear,
  title={Nonlinear backstepping control of grid-forming converters in presence of grid-following converters and synchronous generators},
  author={Karunaratne, Lilan and Chaudhuri, Nilanjan Ray and Yogarathnam, Amirthagunaraj and Yue, Meng},
  journal={IEEE Transactions on Power Systems},
  volume={39},
  number={1},
  pages={1948--1964},
  year={2023},
  publisher={IEEE}
}

@article{samanta2023nonlinear,
  title={Nonlinear model predictive control for droop-based grid forming converters providing fast frequency support},
  author={Samanta, Sayan and Lagoa, Constantino M and Chaudhuri, Nilanjan Ray},
  journal={IEEE Transactions on Power Delivery},
  volume={39},
  number={2},
  pages={790--800},
  year={2023},
  publisher={IEEE}
}

@article{lyu2023grid,
  title={Grid forming fast frequency response for PMSG-based wind turbines},
  author={Lyu, Xue and Gro{\ss}, Dominic},
  journal={IEEE Transactions on Sustainable Energy},
  volume={15},
  number={1},
  pages={23--38},
  year={2023},
  publisher={IEEE}
}

@article{guan2021scheduled,
  title={Scheduled power control and autonomous energy control of grid-connected energy storage system (ESS) with virtual synchronous generator and primary frequency regulation capabilities},
  author={Guan, Minyuan},
  journal={IEEE transactions on power systems},
  volume={37},
  number={2},
  pages={942--954},
  year={2021},
  publisher={IEEE}
}

@article{pawar2021grid,
  title={Grid-forming control for solar PV systems with power reserves},
  author={Pawar, Bandopant and Batzelis, Efstratios I and Chakrabarti, Saikat and Pal, Bikash C},
  journal={IEEE Transactions on Sustainable Energy},
  volume={12},
  number={4},
  pages={1947--1959},
  year={2021},
  publisher={IEEE}
}

@article{ducoin2024analytical,
  title={Analytical design of contributions of grid-forming and grid-following inverters to frequency stability},
  author={Ducoin, Eugenie AS and Gu, Yunjie and Chaudhuri, Balarko and Green, Timothy C},
  journal={IEEE Transactions on Power Systems},
  volume={39},
  number={5},
  pages={6345--6358},
  year={2024},
  publisher={IEEE}
}

@article{dorfler2023control,
  title={Control of low-inertia power systems},
  author={D{\"o}rfler, Florian and Gro{\ss}, Dominic},
  journal={Annual Review of Control, Robotics, and Autonomous Systems},
  volume={6},
  number={1},
  pages={415--445},
  year={2023},
  publisher={Annual Reviews}
}

@article{saha2023impact,
  title={Impact of high penetration of renewable energy sources on grid frequency behaviour},
  author={Saha, Sajeeb and Saleem, MI and Roy, TK},
  journal={International Journal of Electrical Power \& Energy Systems},
  volume={145},
  pages={108701},
  year={2023},
  publisher={Elsevier}
}

@article{shazon2022frequency,
  title={Frequency control challenges and potential countermeasures in future low-inertia power systems: A review},
  author={Shazon, Md Nahid Haque and Jawad, Atik and others},
  journal={Energy Reports},
  volume={8},
  pages={6191--6219},
  year={2022},
  publisher={Elsevier}
}

@article{tomas2025vector,
  title={A vector-based flexible-complexity tool for simulation and small-signal analysis of hybrid AC/DC power systems},
  author={Tom{\'a}s-Mart{\'\i}n, Andr{\'e}s and Zuluaga-R{\'\i}os, Carlos David and Su{\'a}rez-Porras, Jorge and Garc{\'\i}a-Aguilar, Javier and Sigrist, Lukas and Garc{\'\i}a-Cerrada, Aurelio and Kazemtabrizi, Behzad},
  journal={Sustainable Energy, Grids and Networks},
  pages={101817},
  year={2025},
  publisher={Elsevier}
}

@misc{GeneticAlgorithm,
  title = {Genetic {{Algorithm}}},
  urldate = {2025-06-24},
  howpublished = {https://www.mathworks.com/discovery/genetic-algorithm.html},
  langid = {english},
}

@article{girona2024resource,
  title={Resource-aware grid-forming synchronization control: Design, analysis and validation},
  author={Girona-Badia, Jaume and Lacerda, Vin{\'\i}cius Albernaz and Spier, Daniel Westerman and Prieto-Araujo, Eduardo and Gomis-Bellmunt, Oriol},
  journal={IEEE Transactions on Energy Conversion},
  year={2024},
  publisher={IEEE}
}

@article{kenyon2023interactive,
  title={Interactive power to frequency dynamics between grid-forming inverters and synchronous generators in power electronics-dominated power systems},
  author={Kenyon, Rick Wallace and Sajadi, Amirhossein and Bossart, Matt and Hoke, Anderson and Hodge, Bri-Mathias},
  journal={IEEE Systems Journal},
  volume={17},
  number={3},
  pages={3456--3467},
  year={2023},
  publisher={IEEE}
}

@article{peng2022transient,
  title={Transient stabilization control of electric synchronous machine for preventing the collapse of DC-link voltage},
  author={Peng, Yelun and Shuai, Zhikang and Shen, Chao and Hou, Xiaochao and Shen, Z John},
  journal={IEEE Transactions on Smart Grid},
  volume={14},
  number={1},
  pages={82--93},
  year={2022},
  publisher={IEEE}
}

@techreport{ProjectInertiaPhase,
author = {{ENTSO-E}},
year = {2023},
title = {Project {{Inertia}} - {{Phase II}}: {{Updated Frequency Stability Analysis}} in {{Long Term Scenarios}}, {{Relevant Solutions}} and {{Mitigation Measures}}},
  pages = {1--33},
  address = {Rue de Spa, 8, 1000 Brussels, Belgium},
  institution = {ENTSO-E},
  urldate = {2024-06-26},
}

@article{samanta2022fast,
  title={Fast frequency support from grid-forming converters under dc-and ac-side current limits},
  author={Samanta, Sayan and Chaudhuri, Nilanjan Ray and Lagoa, Constantino M},
  journal={IEEE Transactions on Power Systems},
  volume={38},
  number={4},
  pages={3528--3542},
  year={2022},
  publisher={IEEE}
}

@article{samanta2021stability,
  title={Stability analysis of grid-forming converters under dc-side current limitation in primary frequency response regime},
  author={Samanta, Sayan and Chaudhuri, Nilanjan Ray},
  journal={IEEE Transactions on Power Systems},
  volume={37},
  number={4},
  pages={3077--3091},
  year={2021},
  publisher={IEEE}
}

@article{zhao2023small,
  title={Small-signal synchronization stability of grid-forming converters with regulated DC-link dynamics},
  author={Zhao, Liang and Jin, Zheming and Wang, Xiongfei},
  journal={IEEE Transactions on Industrial Electronics},
  volume={70},
  number={12},
  pages={12399--12409},
  year={2023},
  publisher={IEEE}
}

@article{holland1962outline,
  title={Outline for a logical theory of adaptive systems},
  author={Holland, John H},
  journal={Journal of the ACM (JACM)},
  volume={9},
  number={3},
  pages={297--314},
  year={1962},
  publisher={ACM New York, NY, USA}
}

@article{zhou2023analysis,
  title={Analysis of primary frequency regulation characteristics of PV power plant considering communication delay},
  author={Zhou, Wanpeng and Li, Chunlai and Yang, Libin and Li, Zhengxi and Zhang, Chengyun and Zheng, Tianwen},
  journal={Energy Reports},
  volume={9},
  pages={1315--1325},
  year={2023},
  publisher={Elsevier}
}

@article{li2022revisiting,
  title={Revisiting grid-forming and grid-following inverters: A duality theory},
  author={Li, Yitong and Gu, Yunjie and Green, Timothy C},
  journal={IEEE Transactions on Power Systems},
  volume={37},
  number={6},
  pages={4541--4554},
  year={2022},
  publisher={IEEE}
}

@article{mahmood2024evaluating,
  title={Evaluating the Equivalent Inertia of Grid-Following and Grid-Forming Inverter-Based Resources},
  author={Mahmood, Zaid Ibn and Cui, Hantao and She, Buxin and Li, Fangxing Fran},
  journal={IEEE Transactions on Energy Conversion},
  year={2024},
  publisher={IEEE}
}

@article{yang2022internal,
  title={Internal energy based grid-forming control for MMC-HVDC systems with wind farm integration},
  author={Yang, Renxin and Shi, Gang and Zhang, Chen and Li, Gen and Cai, Xu},
  journal={IEEE Transactions on Industry Applications},
  volume={59},
  number={1},
  pages={503--512},
  year={2022},
  publisher={IEEE}
}

@article{gao2025iterative,
  title={Iterative Optimization Method for Frequency Stability Constraints in Renewable Energy-Integrated Power Systems},
  author={Gao, Ruilin and Wang, HF},
  journal={IET Generation, Transmission \& Distribution},
  volume={19},
  number={1},
  pages={e70165},
  year={2025},
  publisher={Wiley Online Library}
}

@article{gothner2021harmonic,
  title={Harmonic virtual impedance design for optimal management of power quality in microgrids},
  author={G{\"o}thner, Fredrik and Rold{\'a}n-P{\'e}rez, Javier and Torres-Olguin, Raymundo E and Midtg{\aa}rd, Ole-Morten},
  journal={IEEE Transactions on Power Electronics},
  volume={36},
  number={9},
  pages={10114--10126},
  year={2021},
  publisher={IEEE}
}
\end{spacing}
\endgroup

\end{document}